\documentclass[a4paper,10pt,twoside]{cpc-hepnp}
\usepackage{CJK,upgreek,fancyhdr}
\usepackage{multicol}
\usepackage{graphicx}
\usepackage{booktabs}
\usepackage{amssymb,bm,mathrsfs,bbm,amscd}
\usepackage[tbtags]{amsmath}
\usepackage{lastpage}

\begin{document}

\fancyhead[c]{\small Chinese Physics C~~~Vol. 37, No. 1 (2013)010201}
\fancyfoot[C]{\small 010201-\thepage}

\footnotetext[0]{Received 31 June 2015}

\title{Empirical pairing gaps and neutron-proton correlations}

\author{%
B.\,S. Ishkhanov,$^{1, 2}$%
\quad S.\,V. Sidorov,$^{1}$%
\quad T.\,Yu. Tretyakova,$^{2;1)}$\email{tretyakova@sinp.msu.ru}%
\quad E.\,V. Vladimirova$^{1, 2}$%
}
\maketitle

\address{%
{$^1$
Faculty of Physics, Lomonosov Moscow State University, Moscow 119991, Russia}\\
{$^2$Skobeltzyn Institute of Nuclear Physics, Lomonosov
Moscow State University, Moscow 119991, Russia}\\
}

\begin{abstract}
Analysis of various mass relations connected with neutron-proton correlations in atomic nuclei is carried out. On the example of $N = Z$ chain it is shown that for self-adjoint nuclei various formulas proposed in literature for $ np $ pairing energy estimations lead to similar results. Significant differences between the calculation methods arise when  nuclei with $N \ne Z$ are considered, which allows to show the complexity of neutron-proton correlations in different types of atomic nuclei and to make some assumptions on the correspondence of a mass ratio to the real effect of $ np $ pairing. The Shell Model parametrization of binding energy makes it possible to arrive to additional conclusions on the structure of mass formulas and their interrelationships with one another.
\end{abstract}



\footnotetext[0]{\hspace*{-3mm}\raisebox{0.3ex}{$\scriptstyle\copyright$}2013
Chinese Physical Society and the Institute of High Energy Physics
of the Chinese Academy of Sciences and the Institute
of Modern Physics of the Chinese Academy of Sciences and IOP Publishing Ltd}%

\begin{multicols}{2}

\section{Introduction}

Over decades since the description of the mechanism of superconducting-type pair correlations  in atomic nuclei \cite{1}, a huge amount of experimental data has been accumulated and a significant number of effective theoretical models has been created that describe the important role of neutron and proton pairs in the formation of various characteristics of atomic nuclei \cite{2,3,4}. However, due to the constant development of experimental capabilities, it became possible to expand the range of the nuclei studied in the region far from stability and  to refine significantly the experimental data on known isotopes, which led to a new wave of theoretical studies of the structure and dynamics of atomic nuclei. One of the important questions  actively discussed  at the present time is the question of neutron-proton correlations in atomic nuclei \cite{WBV06, FM,VI,QW,SBC}. The analysis of $np$ pairing is of particular interest because  it is possible to study the relation between the isoscalar ($T=0$) and isovector ($T = 1$) pairing of nucleons in this case and trace the change of this ratio as a function of the mass number $A$. Traditionally, the main object of the $np$-paring study is a chain of nuclei with $N = Z$. These nuclei demonstrate a vivid example of the isospin symmetry of the nucleon-nucleon interaction, which is a consequence of the charge independence of nuclear forces.

One of the ways to examine the structure of atomic nuclei including the effects of two-nucleons correlations, is a systematic study of the mass surface of atomic nuclei, its global behavior and local fluctuations. This is an important source of information  because experimental values of the nuclear masses are determined with high accuracy and the number of isotopes for which this information is available is increasing constantly \cite{AME2016}. Mass relationships allow one to extract the necessary information on the magnitude of the interaction between nucleons as a function of the mass number $A$ and the occupation probabilities of the subshells near the Fermi energy. For example, it is well known that pairing of identical nucleons leads to stratification of the mass surface and can be quantified from the odd-even staggering (OES) value \cite{GM,BM,Moller_Nix}. Various versions of the estimation of the pairing energy of identical nucleons in even-even isotopes based on the masses of neighbouring nuclei have been studied in detail, but despite the long history of the study of the problem, the question of which relation corresponds to the pair interaction most closely, is still under discussion \cite{JHJ,SDN,BRR,DMN,CQW,CPC}. 

The mass ratios for neutron-proton pairing estimation are more diverse \cite{FM,WCQ,ZCB89,KZ,JC85}. In this case, however, they are mainly considered for nuclei with $N=Z$, and primarily for odd-odd nuclei. These nuclei allows one to address to both isovector spin-zero and isoscalar deutron-like or spin-maximum neutron-proton coupling. Since there are assumptions that  isoscalar pairing of nucleons in heavy nuclei contributes to collective effects significantly, analysis of calculations based on mass ratios should allow one to draw conclusions regarding the effect of the $np$ pairing and the possibility of treating $np$ pairs as deuteron-like states in nuclei. The analisis of mass indicators for chain $N=Z$ is complicated by the presence of Wigner energy, which  is closely connected with $np$-pairing \cite{WBV06, VWB95, SDG97, BF13}.

In the present paper, the ideas underlying various mass relations connected with neutron-proton correlations in different type atomic nuclei are considered. Examples of $N - Z =$~Const chains of nuclei are studied in order to compare the behaviour of indicators under consideration. Binding energy parametrization on the shell model basis makes it possible to clarify the structure of the mass relations obtained and to reveal their interrelation with the $np$ interaction.

\section{Mass relations for $np$-correlations.}\label{MRel}

At present there is a large number of indicators of the $np$ correlations based on the masses of neighbouring nuclei to be found in the literature. Below we consider the basic relations. 

\end{multicols}
\ruleup
\begin{center}
\includegraphics[width=12cm]{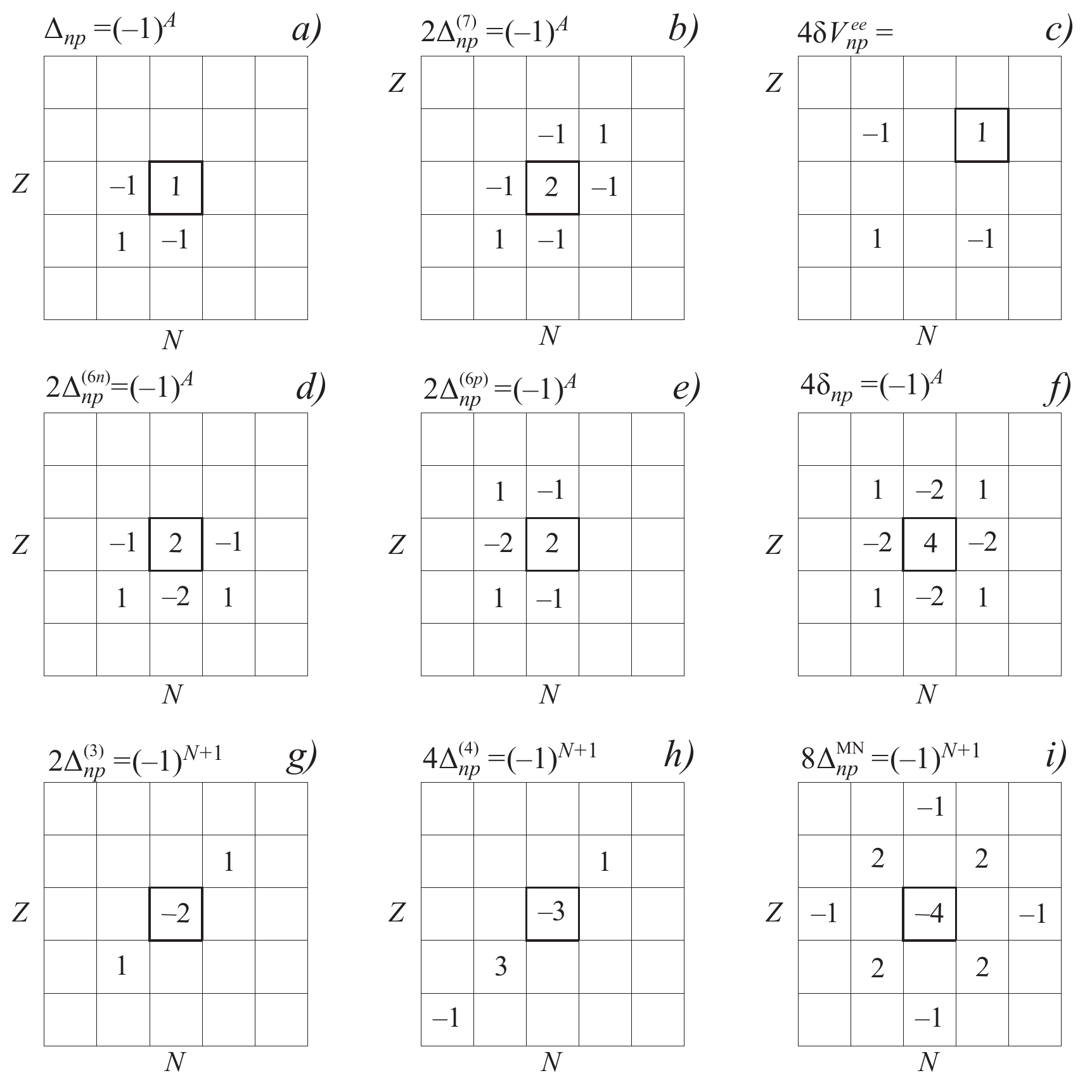}
\figcaption{Diagrams of various indicators of \textit{np}-correlations in nuclei. The coefficients are given for the values of binding energies in the ratios for $a)$ $\Delta_{np}$  -- indicator \eqref{Def_np}, $b)$ $2\Delta_{np}^{(7)}$  --- \eqref{Delta_7}, $c)$ $4\delta V_{np}^{ee}$ --- \eqref{5}, $d)$ $2\Delta_{np}^{(6n)}$ --- \eqref{Delta_6n}, $e)$ $2\Delta_{np}^{(6p)}$ --- \eqref{Delta_6p}, $f)$ $4\delta_{np}$ --- \eqref{dnp}, $g)$ $2\Delta_{np}^{(3)}$ --- \eqref{d3}, $h)$ $4\Delta_{np}^{(4)}$ --- \eqref{d4}, $i)$ $8\Delta_{np}^{MN}$ --- indicator \eqref{D13} for nuclei with even $A$.}
\label{pic: Recipes}
\end{center}

\ruledown
\begin{multicols}{2}

\subsection{Mass relations "from definition" and $\delta V_{np}$ indicator.}\label{Mdef}

In our previous work \cite{CPC} the interrelation of different mass ratios among themselves and their correspondence to the pairing energy of identical nucleons was shown. Various indicators of like nucleon pairing based on the odd-even splitting of the mass surface with different degrees of averaging were considered, and  correspondence of these relations to the explicit definition of the nucleon pairing energy as the difference between two-nucleon separation energy  in nucleus $(A)$ and the doubled one-nucleon separation energy in nucleus $(A-1)$:
\begin{align}
\Delta_{nn}(N,Z) &= S_{2n}(N,Z) - 2S_{n}(N-1,Z), 
\label{Def_nn}
\end{align}
where $S_{2n}$ and $S_{n}$ are two- and one-neutron separation energies respectively. This relation describes the magnitude of neutron pairing.
A similar relation for the proton pairing energy $\Delta_{pp}(N,Z)$ through the proton separation energies $S_{2p}$ and $S_{p}$ can be obtained by swapping $N$ and $Z$. 

To determine the neutron-proton pairing energy in an odd-odd nucleus having an $np$-pair above the double-closed core, one should consider the difference between the $np$ separation energy in  $(N,Z)$ nucleus and the separation energies of a single neutron and a proton in  nuclei $(N,Z-1)$ and $(N-1,Z)$ respectively \cite{Kr59}:
\begin{align}
\nonumber &\Delta_{np}(N,Z)= \\
\nonumber = &S_{np}(N,Z) - [S_n(N,Z-1) + S_p(N-1,Z)] =\\
\nonumber = &B(N,Z) + B(N-1,Z-1) - \\
 &- B(N-1,Z) - B(N,Z-1),
\label{Def_np}
\end{align}
where $S_{np}(N,Z)$ is the $np$-pair separation energy, and $B(N,Z)$ is the binding energy. 
This relation, suggested in \cite{BB} for both even and odd $N$ and $Z$, was widely applied \cite{J72,JB74,MS77,CZA,Fu11,Bul,Fu13,Lu14}.

\begin{center}
\includegraphics[width=7.6cm]{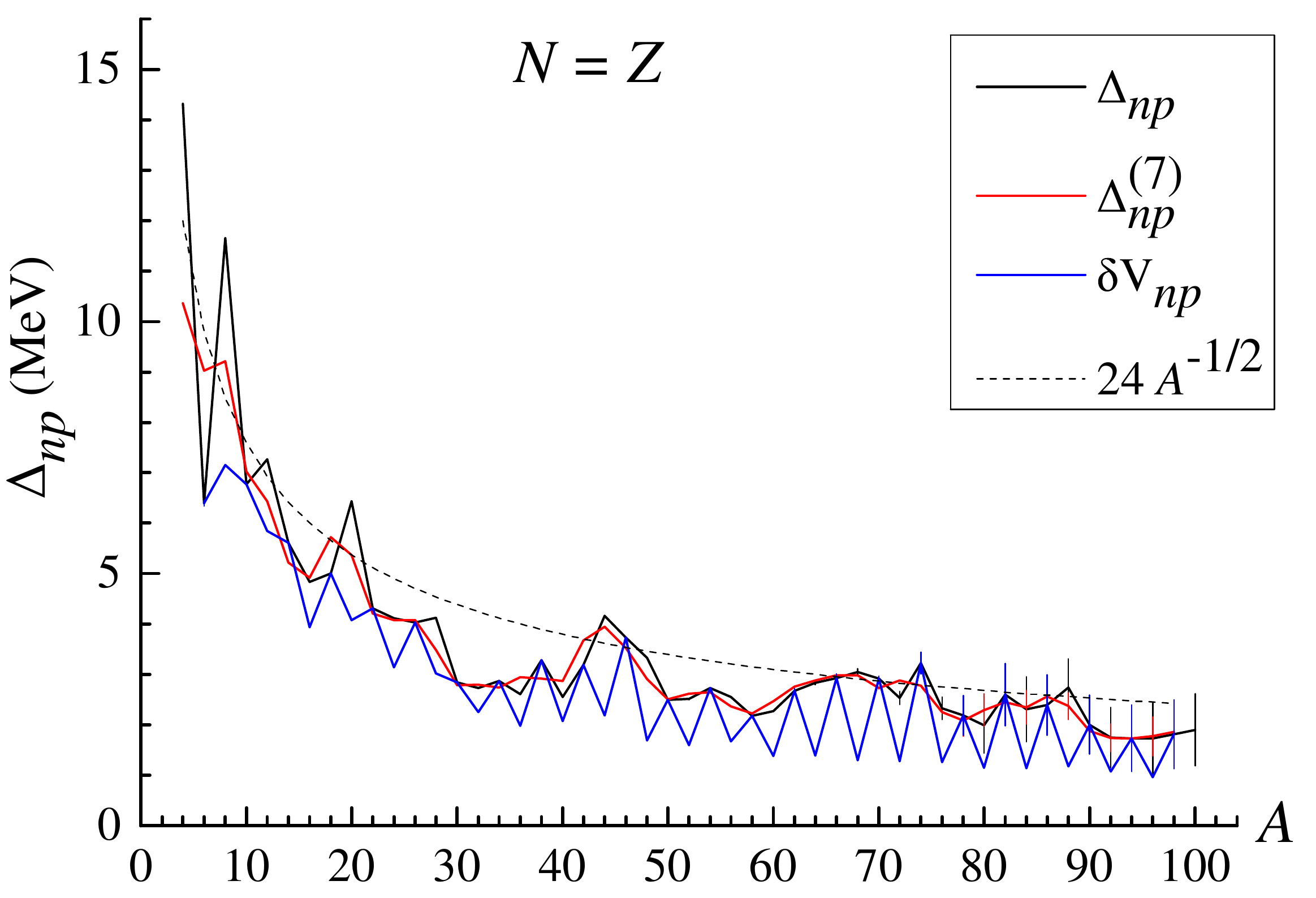}
\figcaption{Indicators  $\Delta_{np}(A)$, $\Delta_{np}^{(7)}(A)$ and $\delta V_{np}$ for even-even and odd-odd nuclei depending on the mass number $ A $ in the nuclei $N=Z$. The dashed line corresponds to $24/A^{1/2}$ dependence. Data on the nuclear masses are from \cite{AME2016}.}
\label{pic: dVnp}
\end{center}

Averaging $\Delta_{np}$ over the nuclei $(N,Z)$ and $(N+1,Z+1)$ belonging to the chain $N-Z=$~Const
\begin{align}
\Delta_{np}^{(7)}(N,Z) = \frac{1}{2}  \left(\Delta_{np}(N,Z) + \Delta_{np}(N+1,Z+1)\right).
\label{Delta_7}
\end{align}
can claim a more accurate estimate of $np$-correlations. Illustrative diagrams with coefficients at  binding energies of neighbouring nuclei in formulas (\ref{Def_np}) and (\ref{Delta_7}) are shown in Fig.~\ref{pic: Recipes}~\textit{a},~\textit{b}. 

Fig.~\ref{pic: dVnp} shows the dependencies of indicators $\Delta_{np}$, $\Delta^{(7)}_{np}$  on the mass number  $A$ in self-adjoint nuclei $N=Z$. The monotonous behaviour of the dependencies and sufficient agreement between the values of $ \Delta_{np}$ and $\Delta^{(7)} _{np}$ for $A>10$ are worth pointing out. Indeed, the results of formula (\ref{Def_np}) for neighbour even-even nuclei and for odd-odd nuclei are very close to each other not only for chain with $N=Z$, but for other isotope regions too. This can be seen from the diagrams in Fig.~\ref{pic: Recipes}: the  difference in the indicators $\Delta_{np}$ for the nuclei $(N,Z)$ and $(N+1,Z+1)$ leads to the well-known Garvey-Kelson mass relations \cite{GK1,GK}:
\begin{align}
\nonumber M(N+2,Z-2) - M(N,Z)+&\\
\nonumber +M(N,Z-1)- M(N+1,Z-2)+&\\
\nonumber +M(N+1,Z)-M(N+2,Z-1)&=0; \\
\nonumber M(N+2,Z) - M(N,Z-2)+&\\
\nonumber +M(N+1,Z-2)- M(N+2,Z-1)+&\\
 +M(N,Z-1)-M(N+1,Z)&=0. 
 \label{GK}
\end{align}
 The accuracy of Garvey-Kelson mass relations is verified on a large number of experimental data and these relationships, like the generalized formulas based on them, are widely used to estimate the mass of nuclei far from stability \cite{Bao13, CZA14}. 

Formally, the proximity of $\Delta_{np}$ values for $o-o$ and $e-e$ nuclei does not necessarily mean that in both cases the indicator displays exactly the $np$-correlations, especially for nuclei with $N=Z$, where the presence of the Wigner cusp significantly changes the picture. For even-even nuclei the applicability of formula (\ref{Def_np}) to estimate the energy of $np$-pairing is not so obvious. 
Indeed, in case of an even number of external nucleons of the same type over a closed core, in addition to  $np$-interaction, like nucleon correlations should also be taken into account. 
Thus, for an even-even nucleus with two $ np$ pairs, $np$-pairing should be defined as the difference between the separation energy of all four nucleons from the core and the separation energies of neutron and proton pairs in nuclei $ (N, Z-2) $ and $ (N-2, Z) $ respectively \cite{VI}:
\begin{align}
\nonumber  \Delta^{ee}_{np}(N,Z) =& \frac14 \left(B(N-2,Z-2) + B(N,Z)-\right.\\
 &\left.-  B(N-2,Z) - B(N,Z-2)\right).
 \label{Def_ee}
\end{align}
The coefficient $1/4$ arises as a result of taking into account the interaction of each proton with each neutron. The corresponding diagram is shown in Fig.~\ref{pic: Recipes}\textit{c}. From diagrams it is seen that for even-even nuclei this indicator may be composed from 4$\Delta_{np}$: 
\begin{align}
\nonumber \delta V_{np}=&\Delta_{np}(N,Z)+\Delta_{np}(N-1,Z-1)\\
\nonumber&-\Delta_{np}(N,Z-1)-\Delta_{np}(N-1,Z).
\end{align}

Difference between the binding energies of four even-even nuclei as an estimation of $np$-interaction energy was proposed in \cite{ZCB89} and analysed  in \cite{BW90,VWB95,SDG97} in connection with structure of Wigner term.  However, in \cite{ZCB89} the indicator $\delta V_{np}$ calculated by (\ref{Def_ee}) was set in accordance to $np$ interaction in odd-odd nuclei $(N+1, Z+1)$ only.  

The indicator $\delta V_{np}$ in both interpretations is still subject to extensive consideration \cite{CC06,BC06,OC06,SC07,Ch09,BH11,WCQ} . In \cite{VWB95} some variant of generalising the formula (\ref{Def_ee}) for different types of nuclei was proposed:
\end{multicols}
\ruleup
\begin{align}
\delta V_{np} (N,Z) =
\begin{cases}
\frac14 [B(N,Z) - B(N,Z-2) - B(N-2,Z) + B(N-2,Z-2)], &  \mbox{ (even, even),}\\
\frac{1}{2}[B(N,Z) - B(N,Z-1) - B(N-2,Z) + B(N-2,Z-1)], & \mbox{  (even,  odd),}\\
\frac{1}{2}[B(N,Z) - B(N,Z-2) - B(N-1,Z) + B(N-1,Z-2)], &  \mbox{  (odd,   even),}\\
B(N,Z) - B(N,Z-1) - B(N-1,Z) + B(N-1,Z-1), &  \mbox{  (odd,   odd).}\\
\end{cases}
 \label{5}
\end{align}
\vspace {20pt}

\ruledown

\begin{multicols}{2}

In Fig.~\ref{pic: dVnp} the last variant of indicator $\delta V_{np}(A) $ in comparison with $\Delta_{np}$ is presented. In this case indicators $\Delta_{np}$ and $\delta V_{np} $ coincide for odd-odd nuclei, formula (\ref{5}) for even-even nuclei produces consistently lower estimates of $np$ pairing energy.  The dependence of $\delta V_{np}$ acquires a pronounced zigzag character due to the relation $\delta V_{np}^{oo}> \delta V_{np}^{ee}$. Since  $\delta V_{np}(A) $ empirically obtained, it may contain a different components of different nature. The chain $N=Z$ is an anomalous case due to Wigner energy. On the other hand the structure of Wigner term connected with $np$-pairing and considerations of empirical relations for  $\delta V_{np}$ can help to clarify it \cite{VWB95}. 

\subsection{Wigner term}\label{Wigner}

For the purpose to consider  $np$-correlation the so-called Wigner term is of special importance. This contribution was firstly considered on the basis of analysis of the SU(4) spin-isospin symmetry of nuclear forces by Wigner \cite{W37}, who showed that the symmetry energy in addition to a term, proportional $(N-Z)^2/A$, must also has  a contribution proportional to isospin asymmetry $|I|$ ($I =(N-Z)/A$), which leads to enhancement of binding energy   near $N=Z$. In mass formula of droplet model Wigner term was adopted in form  \cite{M76}
\begin{align}
\nonumber E_W &= W(|I| +d),\mbox{  where }W=30\mbox{ MeV}, \\
 \nonumber d&=\begin{cases}
\frac{1}{A}  \mbox{ (odd-odd), }N=Z\\
0 \mbox{  otherwise.}\\
\end{cases}
\end{align}
The correction for ($N=Z$) odd-odd nuclei $d$-term  was added "because it clearly called for by the experimental masses (see \cite{GK}, Table I)".   The generalization of Wigner term, performed in \cite{JHJ}, results in three terms:
 \begin{align}
\nonumber E_W &= -b_1|I|+b_2/a+b_3/A, 
 \end{align}
where $b_3$-term corresponds to ($N=Z$) odd-odd nuclei and  $b_2$-term connected with possible $\alpha$-correlation effect. Currently the most common expression for Wigner term is
\begin{align}
 E_W &= W(A)|N-Z|+d(A)\pi_{np}\delta_{NZ}, 
  \label{Wign}
 \end{align}
where $\pi_{np}=\frac14(1-\pi_n)(1-\pi_p)$, $\pi_n=(-1)^N$ and $\pi_p = (-1)^Z$ being the nucleon-number parities. The question about $d/W$ is still open: as mentioned above  some estimates suggest that ratio $d/W = 1$ \cite{M76}, analysis of experimental masses leads to $d/W = 0.56 \pm 0.27$ \cite{JHJ}. It seems productive to use empirical mass relations for definition of Wigner term parameters. Indicator $\delta V_{np}$ (\ref{Def_ee}) was used for investigating the $np$-correlation energy  and it was shown, that it sensitive for Wigner energy and can be used as $d$-term in expression (\ref{Wign}) \cite{BW90, VWB95}. The mass relations for $\delta V_{np}$, obtained in   \cite{VWB95} form supermultiplet theory were given above (see(\ref{5})). In \cite{SDG97} a certain combinations of $\delta V_{np}(N,Z)$ were suggested to define $W(A)$ and $d(A)$. The difference between even-even and odd-odd nuclei is not limited to the presence of a special $d$-term, the mass relation for $W(A)$ is also different in these two cases:
\begin{align}
\nonumber \mbox{ for }N=&Z,\mbox{ even-even}\\
\nonumber W(A)=&\delta V_{np}(N,Z)-\\
&-\frac12[\delta V_{np}(N,Z-2)+\delta V_{np}(N+2,Z)]\\
\nonumber \mbox{ for }N=&Z,\mbox{ odd-odd}\\
\nonumber W(A)=&-\delta V_{np}(N+1,Z-1)+\\
&+\frac12[\delta V_{np}(N-1,Z-1)+\delta V_{np}(N+1,Z+1)]\\
\nonumber d(A)=&-4\delta V_{np}(N+2,Z)+\\
&+2[\delta V_{np}(N+1,Z-1)+\delta V_{np}(N+3,Z+1)]. \label{W_d}
\end{align}
Experimental values consistent with the simple relation $d_{T=0}/W \approx 1$. Analysis of the Wigner energy in terms of $np$ pairs of a given angular momentum and isospin shows that the Wigner term cannot be solely explained in terms of correlations deutron-like $np$ pairs.

There are still a lot of efforts to determine the precise structure of symmetry energy  and to extract the Wigner term \cite{WBV06, MS97, BF13, V00, TC16}. The $d-$term interpretation based on mass relation gives rise to discussions in literature. It seems to be useful to consider different mass ratios for $np$-correlations not only for odd-odd $(N=Z)$ nuclei,  but for nuclei with different $N$, $Z$ parity with $N-Z \ge 1$ too.

\subsection{$np$-correlation from $S_n$ and $S_p$}\label{nn_pp}

Estimates of  $np$-correlations can be obtained by consideration of either neutron or proton separation energies along the chains of isotones or isotopes respectively.  Indeed, it follows from (\ref{Def_np}) that for odd-odd nuclei
\begin{align}
\nonumber \Delta_{np}(N,Z) &=  [S_n(N,Z) - S_n(N,Z-1)]=\\
 \nonumber &= [S_p(N,Z) - S_p(N-1,Z)].
\end{align}

Fig.~\ref{pic: Sn(N)} shows the separation energies of the neutron $S_n$ and the proton $S_p$ in  isotopes Sn ($Z = 50$) and Sb ($Z = 51$) as functions of the number of neutrons. 

\begin{center}
\includegraphics[width=7.6cm]{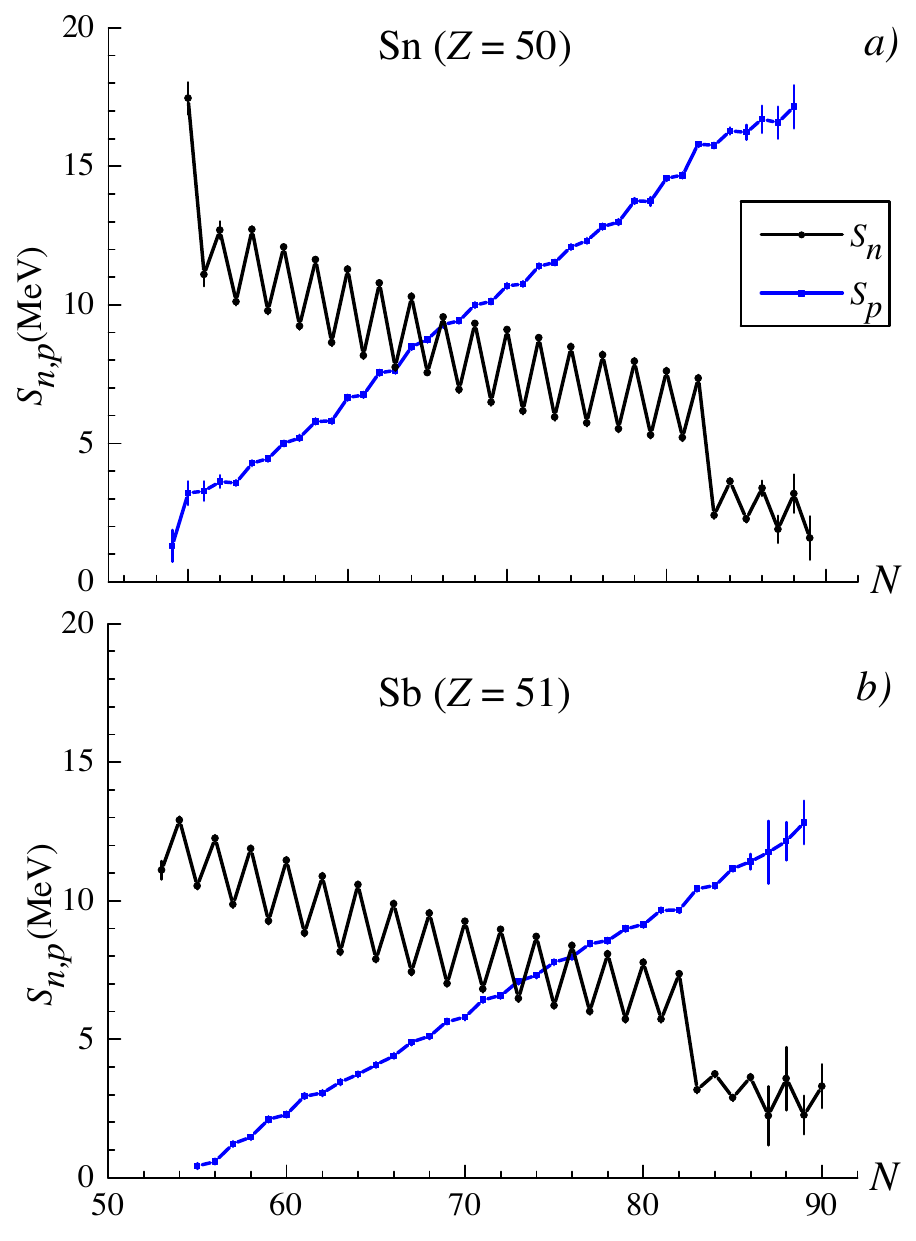}
\figcaption{Proton and neutron separation energies $S_p$ and $S_n$ in Sn (a) and Sb (b) isotopes.  Data on the nuclear masses are taken from \cite{AME2016}.}
\label{pic: Sn(N)}
\end{center}

The dependence of $S_n(N)$ has a zigzag character, which is related to neutron pairing.  At the same time, the dependence of $S_p(N)$  demonstrates even-odd jumps as well despite the constancy of $Z$, due to additional interaction of a proton with an odd neutron. The distance between parallel lines drawn through isotopes with even and odd $Z$ must correspond to  $np$-interaction  \cite{GS55,Sh57}.

Schematically, the dependence  $S_n(Z)$ in a chain of isotones is shown in Fig.~\ref{pic: D6p} (b). Different behaviour of the dependencies for the even and odd number of neutrons is of importance: in case of an even $N$, the largest values of $S_n$ also correspond to even values of $Z$; for isotones with odd $N$, the maxima correspond to odd values of $Z$. This feature of the $S_n(Z) $ and $ S_p (Z)$ dependencies was explained  in \cite{Sh57} in the frame of the shell model. According to the scheme in Fig.~\ref{pic: D6p} (b), the expression for $\Delta_{np}$ should include the dependence on parity of $A$:
\begin{align}
\nonumber \Delta_{np}(N,Z) &= (-1)^A [S_n(N,Z) - S_n(N,Z-1)]=\\
 &= (-1)^A [S_p(N,Z) - S_p(N-1,Z)].
\label{Def_np11}
\end{align}
\end{multicols}
\begin{center}
\includegraphics[width=12cm]{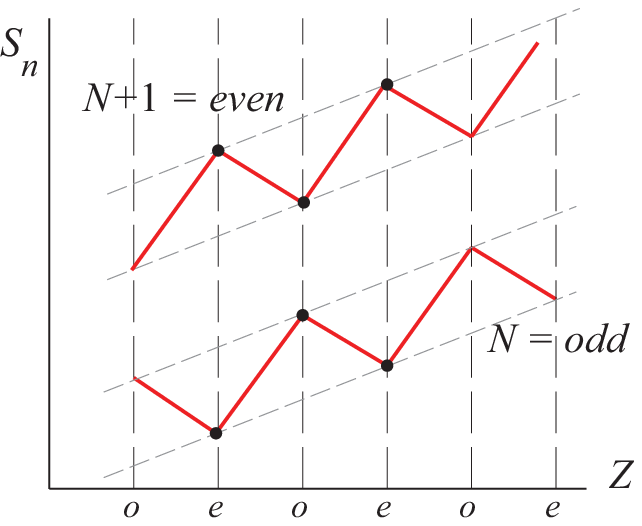}
\figcaption{ Schemes used to determine the characteristics of pairing interaction  from the nucleon separation energies. $a)$ like-nucleon correlation ($S_n(N)$ for $Z=$~Const), $b)$ neutron-proton correlation ($S_n(N)$ for $N=$~Const)}
\label{pic: D6p}
\end{center}

\ruledown
\begin{multicols}{2}

However, the experimental dependencies $S_p(Z)$  depicted in Fig.~\ref{pic: Sn(N)}  show that the problem of introducing a dependence on parity of $A$ is not so obvious, since the relations $S_p^{oo}(N,Z)> S_p^{eo}(N+1,Z)$ and $S_p^{ee}(N+1,Z)> S_p^{oe}(N+1,Z+1)$ are not always satisfied. In fact, these inequalities appear to be invalid in most cases. Therefore, when constructing experimental dependencies for $\Delta_{np}$, a dependence on  parity of $A$  is not taken into account \cite{Z58,VI18}. Since the study of empirical values of $np$-pairing is based on the chains of nuclei with even $A$, the question of  dependence on $A$-parity is not so significant. Further more, the value of $\Delta_{np}$ in odd-$A$ nuclei is close to zero. None the less, we choose to keep the $(-1)^A$ 'phase' from the mass ratio construction point of view.

 As in the case of relations for identical nucleon pairing \cite{CPC}, it seems reasonable to use  the values of $\Delta_ {np} $ averaging for two or more neighbouring nuclei, which leads to  formulas \cite{JHJ}:
\begin{align}
\nonumber \Delta_{np}^{(6,n)}(N,Z) = &\frac{1}{2} [\Delta_{np}(N+1,Z) +\Delta_{np}(N,Z)]=\\
\nonumber= &\frac{ (-1)^A}{2} [-S_n(N+1,Z) - S_n(N,Z-1) +\\
&+ S_n(N,Z) + S_n(N+1,Z-1)]. \label{Delta_6n}
\end{align}

Similar consideration of the scheme for the proton separation energy $S_p$ in  isotones $ Z=$~Const leads to the formula:
\begin{align}
\nonumber \Delta_{np}^{(6,p)}(N,Z) =& \frac{1}{2} [\Delta_{np}(N,Z+1) + \Delta_{np}(N,Z)]=\\
\nonumber = &\frac{ (-1)^A}{2} [-S_p(N,Z+1) - S_p(N-1,Z) +\\
&+ S_p(N,Z) + S_p(N-1,Z+1)]. \label{Delta_6p}
\end{align}

In Fig.~\ref{pic: Recipes}~\textit{d},~\textit{e}  diagrams  for  \eqref{Delta_6n} and \eqref{Delta_6p} indicators are shown. 
As can be seen from the relations above, \eqref {Delta_6n} serves as averaging of the two differences  of neutron separation energies for even and odd $A$ using two chains of isotones: $N$ and $N + 1$.    The formula that averages the neutron separation energy  differences both in neighbouring isotonic chains and for neighbouring nuclei in each chain $Z$ and $Z + 1$ (see Fig.~\ref{pic: Recipes}\textit{f}) is the most symmetrical.  This scheme shows that averaging in accordance with formulas \eqref {Delta_6n} and \eqref {Delta_6p} leads to the same estimates. Indeed, it is evident from the diagrams in the fig.~\ref{pic: Recipes}~\textit{d},~\textit{e} that the difference in these values brings us to well-known Garvey-Kelson mass relations (\ref{GK}).
\begin{center}
\centering
\includegraphics[width=8cm]{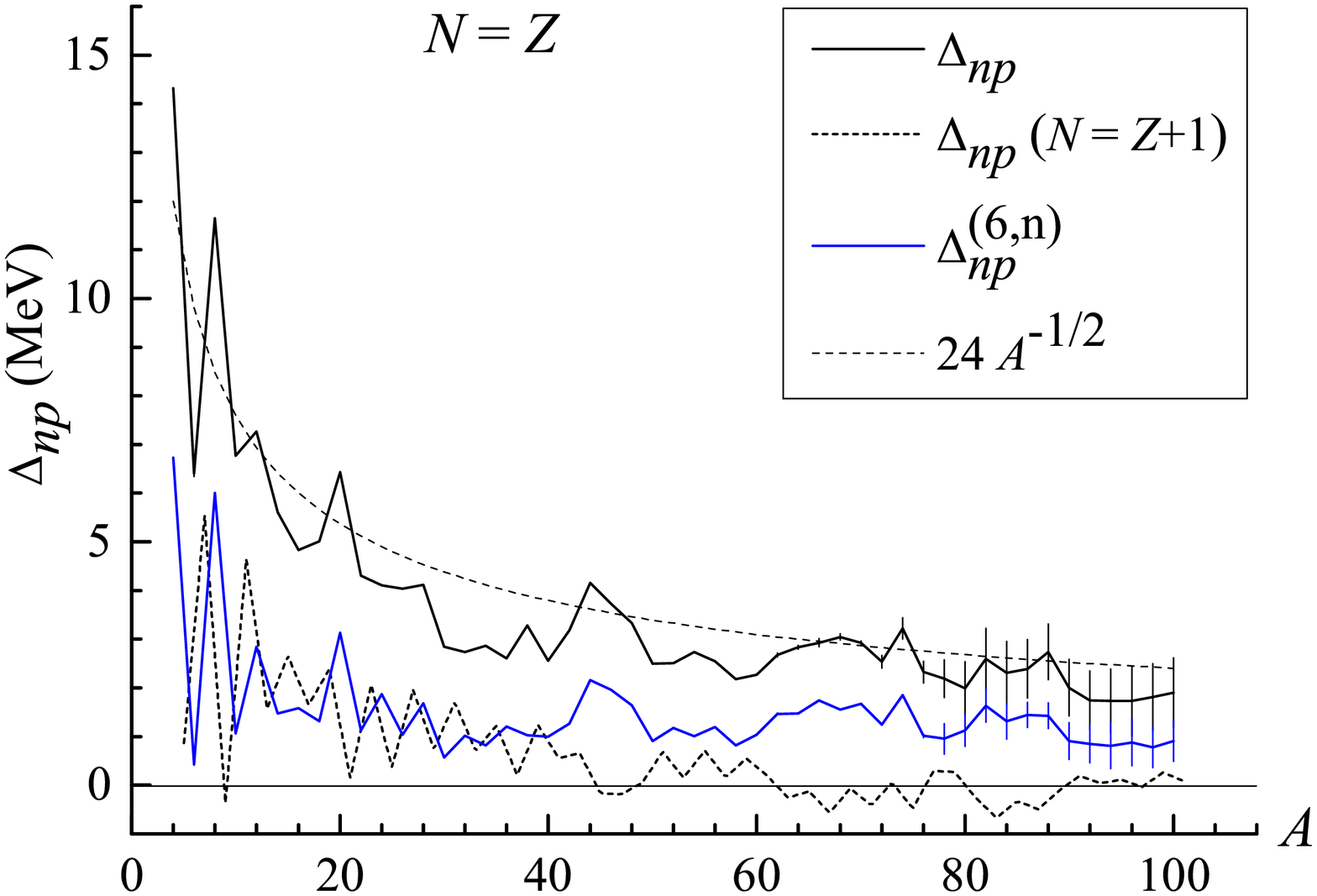}
\figcaption{Indicators $\Delta_{np}$ ( black solid line) and $\Delta_{np}^{(6,n)}$  (blue) in nuclei with $N=Z$. (Dotted line shows the dependence of $\Delta_ {np} (A) $ in nuclei $N= Z + 1$, dashed line corresponds to $24A^{-1/2}$).}. 
\label{pic: D6}
\end{center}

Fig.~\ref{pic: D6} presents the indicators $\Delta_{np}(A)$ for a chain of nuclei with $N = Z$ consisting of even-even and odd-odd isotopes, and for a neighbouring chain of  odd nuclei with $N=Z + 1$. While the values $\Delta_{np}$ for $N = Z$ are high and in general are in accordance with the analytical ratio $24A^{- 1/2}$, the corresponding values for odd nuclei ($N=Z+1$ chain) are essentially smaller and  tail to zero, acquiring even negative values for higher values of $A$. Accordingly, for the chain with $N = Z$ indicator $\Delta_{np}^{(6,n)} (A) $ (\ref{Delta_6n}), which is the averaging between these two chains, lays substantially below $\Delta_{np} $. This example illustrates the contribution of symmetry energy for $N = Z$ nuclei the best, although this trend holds for nuclei with neutron excess too.

Indicator $\Delta_{np}^{(6,n)} (A) $ (\ref{Delta_6n}) was used as an estimation of $np$-pairig term in several papers  \cite{JHJ, WCQ, CQW} . These indicators include difference between nuclei with different $A$-parity, therefore they reflect the complexity of $np$-correlations, not only the $np$-pairing in odd-odd nuclei. Thus, this relationship can be interpreted differently, for example as an indication of $\alpha$-clustering effects in even-even nuclei \cite{ZG88}.

\subsection{Mass relations based on deuteron separation energy.}\label{Deu}

The indicators of the $np$-pairing $\Delta_{np}$ (\ref {Def_np}) and $\Delta^{(7)}_{np}$ (\ref{Delta_7}) are determined by the masses of neighbouring nuclei with both even and odd $A$, as well as $N$ and $Z$. Because the corresponding estimates of like nucleon pairing are based on isotone or isotope chains of nuclei, a significant difference is seen between these estimates and indicators (\ref {Def_np}), (\ref{Delta_7}).

\begin{center}
\includegraphics[width=8cm]{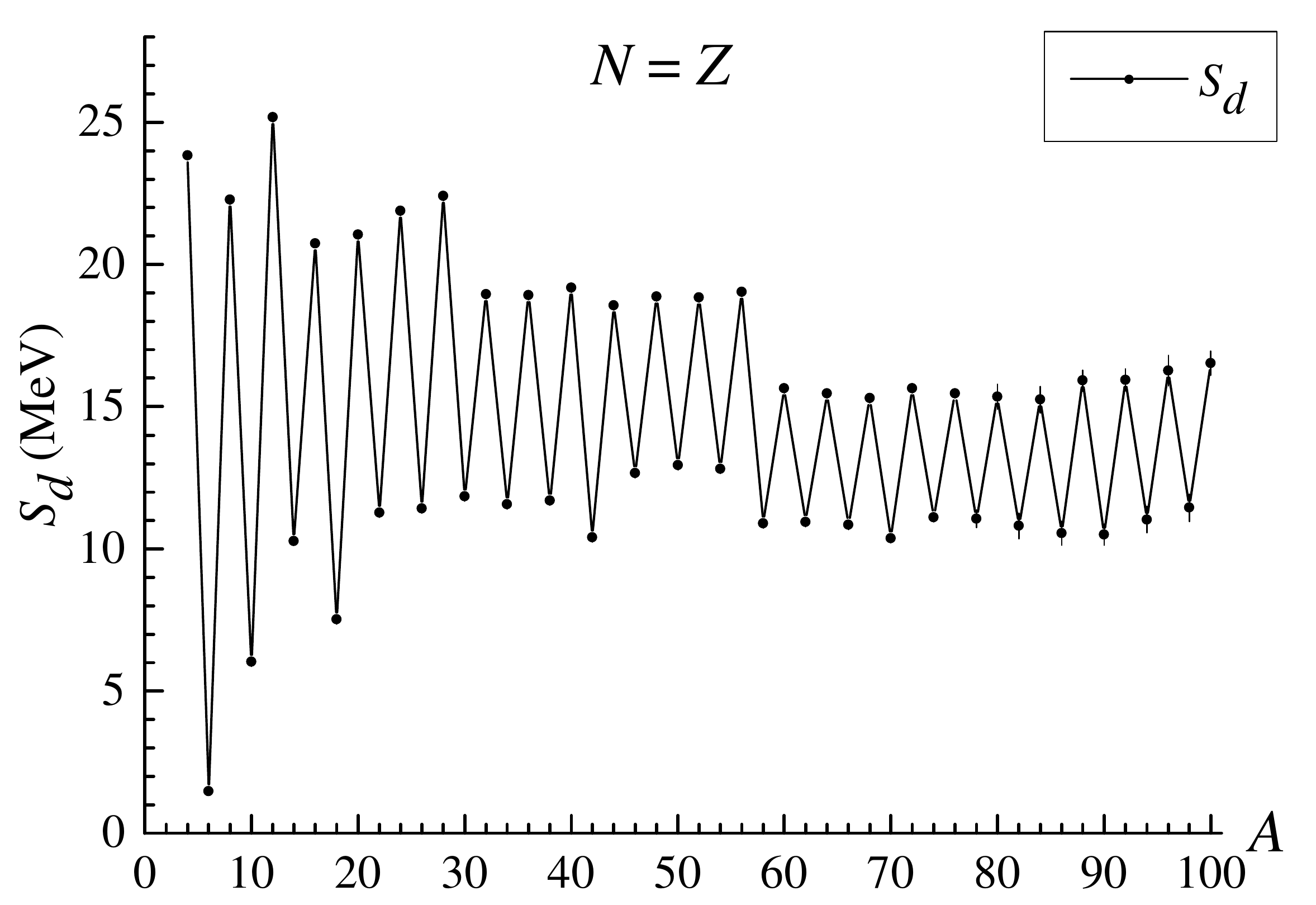}
\figcaption{Deuteron separation energy $S_d(A)$ in nuclei with $N=Z$. Data on the nuclear masses are taken from \cite{AME2016}.}
\label{deutron}
\end{center}

Variants of $np$-pairing indicators constructed by analogy with the formulas for neutron OES and proton OES calculation, use the binding energies of even or odd $A$ nuclei along the chain with $ N-Z=$~Const. Indeed, for a chain of even-$A$ nuclei, one notes the splitting of binding energies into two groups for even-even and odd-odd nuclei. Since the mass number $A$ grows quickly for this sequence, the splitting is too small against the background of a general increase in  binding energy. As in the case of like nucleons, this effect is more pronounced for the difference in binding energies of neighbouring isotopes \cite{CPC}. In the case of the $N=Z$ chain, this corresponds to the deuteron separation energy, corrected for its binding energy:
$$S_d(N,Z) = B(N,Z) - B(N-1,Z-1)-2.22\mbox{ MeV}$$

Fig.~\ref {deutron} depicts the $A$-dependence of the deuteron separation energy $S_d$ in nuclei with $N = Z$. Like $S_n(Z)$ and $S_p(N)$, it shows a zigzag character with an overall tendency to a gradual decrease and stabilization of the even-even -- odd-odd splitting for heavier isotopes. The energy of $np$-pairing in odd-odd nucleus ($N, Z$) based on this dependence corresponds to half the difference in the deuteron separation energies for the even-even and odd-odd nuclei:
\begin{align}
\nonumber \Delta_{np}^{(3)}(N,Z) =&\\
\nonumber=&\frac{1}{2}\left(S_d(N+1, Z+1) - S_d(N,Z)\right) =\\
\nonumber = & \frac{1}{2}\left(B(N+1,Z+1)-\right.\\
& \left.- 2B(N,Z)+B(N-1,Z-1)\right).
\label{d3}
\end{align}

This relation was used in \cite{MF00, BF13} for estimations of isovector $np$-interaction. Indeed, one can see that in the case of even $A$  deuteron separation energy is no more than distance between even-even and odd-odd mass surfaces, corrected on deuteron binding energy $B_d$. The averaging indicator $\Delta^{(3)}_{np}$ cancels the $B_d$ and correspond to 
$$BE_{ee}-BE_{oo} \approx \Delta_p + \Delta_n \approx 2\Delta.$$
 Charge independence of nuclear forces leads to the fact, that isovector $np$-pairing in odd-odd $(N=Z)$ nuclei must be the same as neutron pairing in neighbouring  $(N+1,Z-1)$ isotope   and proton pairing in $(N-1,Z+1)$ isotope. So indicator $\Delta^{(3)}_{np}$ in the chain of $N=Z$ isotopes can be used  for $np$-correlation study. It must be different for isotopes with $N-Z \ge 2$, but nevertheless it makes sense to trace the behaviour of indicators constructed by analogy with mass ratios for like-nucleon pairing for isotopes chains with $N-Z=\mbox{ Const }\ge 2$.
 
 \begin{center}
\includegraphics[width=8cm]{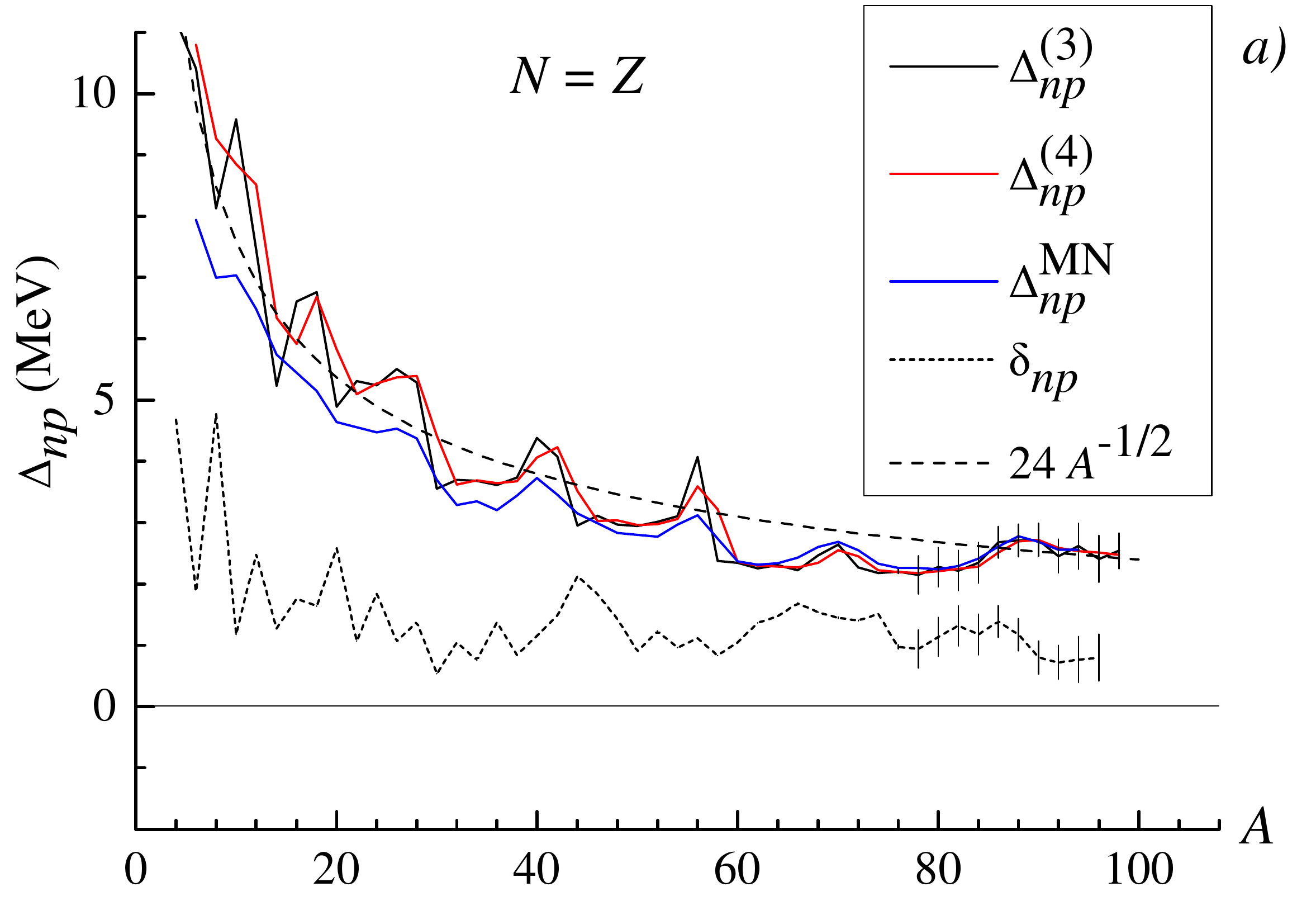}
\includegraphics[width=8cm]{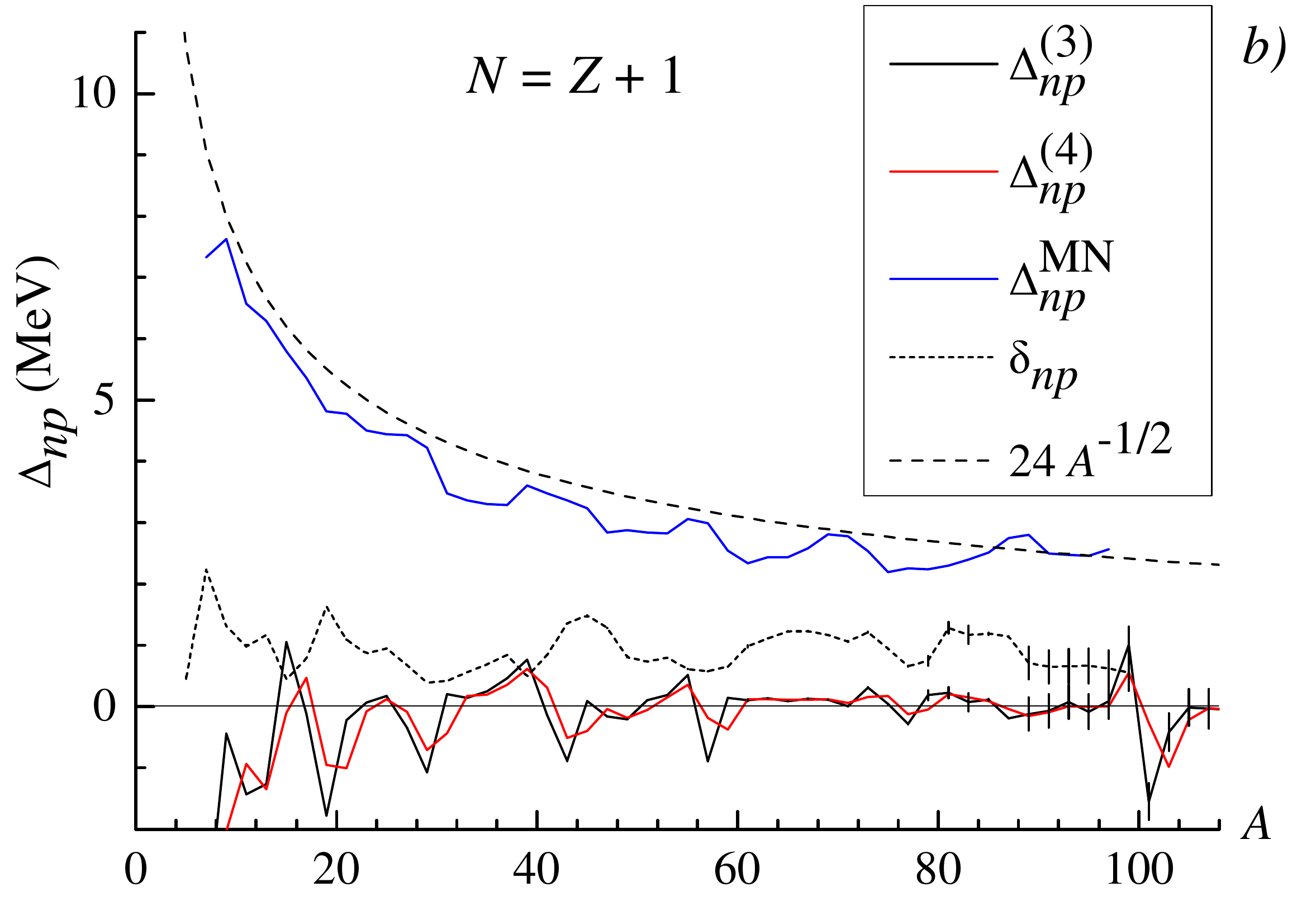}
\includegraphics[width=8cm]{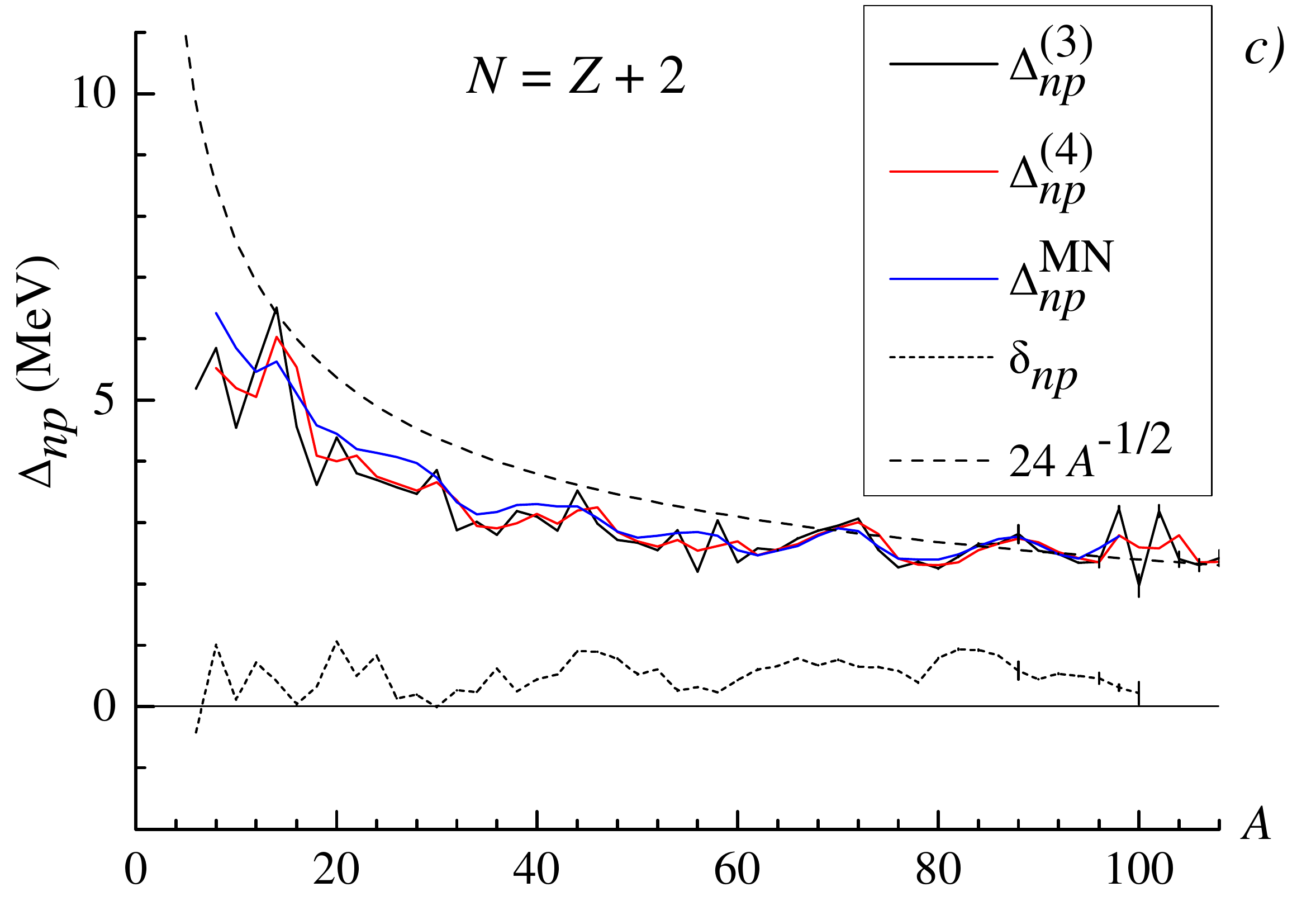}
\figcaption{Indicators of $np$-correlations in  the chains a) $N=Z$, b) $N=Z+1$, c) $N=Z+2$: $\Delta_{np}^{(3)}(A)$ (red line),  $\Delta_{np}^{(4)}(A)$ (green line),   $\Delta_{np}^{(13)}(A)$ (blue line) ,    $\delta_{np}(A)$ (dashed black line).  The thin dotted line corresponds to $24/A^{1/2}$ dependence. Data on the nuclear masses are taken from \cite{AME2016}.}
\label{pic:dnp}
\end{center}

The relation (\ref{d3}) is analogous to the formula for OES related to neutron pairing \cite{BM}:   
\begin{align}
\nonumber \Delta_{n}^{(3)}(N,Z)= &\\
\nonumber=&\frac{(-1)^{N+1}}{2}\left(S_n(N+1, Z) - S_n(N,Z)\right) =\\
\nonumber= &\frac{(-1)^{N+1}}{2}\left(B(N+1,Z) -\right.\\
&\left.- 2B(N,Z)+B(N-1,Z)\right).
\label{dnn3}
\end{align}

 By analogy with the averaged estimates of the OES effect, one can introduce an indicator based on binding energies of four nuclei \cite{Audi}:
\begin{align}
\nonumber \Delta_{np}^{(4)}(N,Z)= &\\
\nonumber = &\frac{1}{2}\left(\Delta_{np}^{(3)}(N, Z) + \Delta_{np}^{(3)}(N-1,Z-1)\right) =\\
\nonumber & = \frac{(-1)^{N+1}}{4}\left(S_d(N+1,Z+1) -\right. \\
& \left.-2S_d(N,Z) + S_d(N-1,Z-1)\right).
\label{d4}
\end{align}
The diagrams of the coefficients for  $\Delta_{np}^{(3)}$ and $\Delta_{np}^{(4)}$ calculation are given in Fig.~\ref{pic: Recipes}~\textit {g}, ~\textit {h} respectively. The $(-1)^{N+1}$ multiplier is used for chains of even-$A$ nuclei only. For these chains, OES effect appears to be prominent, with deuteron separation energy of even-even nuclei being consistently greater than that of odd-odd nuclei. No such relation takes place for odd-A nuclei, and so the $(-1)^{N+1}$ factor is ommited in calculations for the corresponding chains.

The values of indicators $\Delta_{np}^{(3)}$ and $\Delta_{np}^{(4)}$ for the nuclei with $N = Z$  are shown in Fig.~\ref{pic:dnp} a). Since the deuteron separation energy $S_d(A)$ does not have a common slope, the quantities $\Delta_{np}^{(3)}(N, Z)$ and the averaged characteristic $\Delta_{np}^{(4)}(N, Z)$ practically coincide. The dependence has a smooth character with jumps in regions of doubled magic numbers  16, 40, 56. The general course of the dependencies is in accordance with the approximation \cite{BM} $2\Delta = 24/A^{1/2}$; in the region of light nuclei, the majority  of $\Delta_{np}^{(3)} $ and $\Delta_{np}^{(4)} $ values lay above, and for $A>40$ -- below this approximation. Farther on  Fig.~\ref{pic:dnp} are shown  $\Delta_{np}^{(3)}$ and $\Delta_{np}^{(4)}$ versus $A$ plots for chains of odd-$A$ nuclei with $N=Z+1$ (b) and even-$A$ nuclei with $N=Z+2$ (c).
From Fig.~\ref{pic:dnp} (b) it is clear that for most nuclei with odd $A$ the both indicators have practically zero values. These characterestics show similar behaviour (Fig.~\ref {pic:dnp} (a) and (c)), but in the case of $N=Z + 2$,   $\Delta_{np}^{(3)} $ and $ \Delta_{np}^{(4)} $ values are smaller due to absence of  Wigner term.

As we see later,  $\Delta_{np}^{(3)}$ and $\Delta_{np}^{(4)}$ have a complex structure and are inderectly related to $np$-correlations. That is why in different papers they have different interpretations. For example, in \cite{JHJ} the  indicator $\Delta_{np}^{(4)}$ was proposed as four-nucleons correlation estimation.

\subsection{Mass surface OES.}\label{EoS}

The mass surface splitting  is primarily due to the pairing of identical nucleons, but the experimental estimate of fluctuation between the masses of even-even and odd-odd nuclei is somewhat less than the sum of OES effect of  protons $\Delta_p$ and  neutrons $\Delta_n$.
This discrepancy is generally attributed to the presence of residual neutron and proton interactions \cite{BM}, and in order to calculate the splitting between mass surfaces for even-even and odd-odd nuclei one uses the relation \cite{Gupta,MN,S03}:
 \begin{equation}
 E_{ee} - E_{oo} = \Delta_n+\Delta_p-\delta.
 \end{equation}
     Сorrection $\delta $ arising from  residual attractive interaction of the unpaired proton and the unpaired neutron, is interpreted as the value of the $np$ interaction and approximated by the dependence $\delta = 20/A$~MeV  \cite{BM}.
     
Madland and Nix \cite{MN} obtained equations for $\Delta_n$, $\Delta_p$ and $\delta$ in finite differences on the basis of Taylor series expansion to the fourth-order derivatives. Thus, if  values of five neighbouring isotopes or isotones are used to calculate the OES effect for neutrons $\Delta_n$ and protons $\Delta_p$, then data for a substantially greater number of neighbouring nuclei are required to calculate the $np$-interaction indicators:
\begin{equation}
\Delta_{np}^{MN}(N,Z)= \Delta_{n} + \Delta_{p} - \delta_{np}, \label{D13}
\end{equation}
where  $\delta_{np}(N,Z)$ is the $np$-interaction correction:
\end{multicols}
\ruleup
\begin{align}
\nonumber \delta_{np}(N,Z) &= \frac{(-1)^A}{4}\left(2 [B(N+1,Z) + B(N-1,Z) + B(N,Z+1) + B(N,Z-1)] - 4B(N,Z) -\right.\\
&\left.- [B(N+1,Z+1) + B(N-1,Z+1) + B(N-1,Z-1) + B(N+1,Z-1)]\right). \label{dnp}
\end{align}
The protons and neutrons OES in this case depend on the parity of the number of corresponding nucleons: 
\begin{align}
\Delta_{n} =
\begin{cases}
\Delta_{n}^{(5)}(N,Z), &   \mbox{ even } N\\
\Delta_{n}^{(5)}(N,Z)+\delta_{np}, &    \mbox{ odd } N
\end{cases}
\label{eo}
\end{align}
\begin{align}
\Delta_{p} =
\begin{cases}
\Delta_{p}^{(5)}(N,Z), &   \mbox{ even } Z\\
\Delta_{p}^{(5)}(N,Z)+\delta_{np}, &   \mbox{ odd } Z
\end{cases}
\label{oe}
\end{align}
\begin{subequations}
\begin{align}
\Delta_{n}^{(5)}(N,Z) = \frac{(-1)^N}{8} [S_n(N+2,Z) - 3S_n(N+1,Z) +3S_n(N,Z)-S_n(N-1,Z) ],\\
\Delta_{p}^{(5)}(N,Z) = \frac{(-1)^Z}{8} [S_p(N,Z+2) - 3S_p(N,Z+1) +3S_p(N,Z)-S_p(N,Z-1) ].
\end{align}\label{d5}
\end{subequations}



\begin{center}
\tabcaption{\label{Cb} Parameters of the fitting $\Delta_{np}(A) = C\cdot A^{-b}$ in nuclei with $N=Z$ and $N-Z=2$.}
\begin{tabular*}{120mm}{c@{\extracolsep{\fill}}ccccc}
\toprule 
  & $N=Z$ & & $N-Z=2$ & \\
  \hline
  & $C$~(MeV) & $b$  & $C$~(MeV) & $b$\\
\hline
$\Delta_{np}(A)\,$ & $\,\,\,\,29.4 \pm 1.8\,\,\,\,$ & $0.60 \pm 0.02\,\,\,\,$ & $\,\,\,\,5.8 \pm 0.8\,\,\,\,$ & $0.37 \pm 0.04\,\,\,\,$\\

$\Delta_{np}^{(7)}(A)$ & $\,\,\,\,23.3 \pm 1.6\,\,\,\,$ & $0.53 \pm 0.03\,\,\,\,$ & $\,\,\,\,6.0 \pm 0.6\,\,\,\,$ & $0.39 \pm 0.03\,\,\,\,$\\
\hline
$\Delta_{np}^{(6n)}(A)$ & $10.3 \pm 1.4$ & $0.56 \pm 0.05$\,\,\,\,& $\,\,\,\,1.0 \pm 0.3\,\,\,\,$ & $0.14 \pm 0.08\,\,\,\,$\\

$\Delta_{np}^{(6p)}(A)$ & $9.3 \pm 1.3$ &$0.52 \pm 0.04$ \,\,\,\,& $\,\,\,\,0.0 \pm 0.1\,\,\,\,$ & $-0.7 \pm 0.3\,\,\,\,$\\

$\delta_{np}^{}(A)$ & $6.9 \pm 1.1$ & $0.45 \pm 0.05$ & $\,\,\,\,0.2 \pm 0.1\,\,\,\,$ & $0.2 \pm 0.1\,\,\,\,$\\
\hline
$\Delta_{np}^{(3)}(A)$ & $25.9 \pm 1.3$ & $0.53 \pm 0.02$ & $\,\,\,\,10.8 \pm 0.8\,\,\,\,$ & $0.33 \pm 0.02\,\,\,\,$\\

$\Delta_{np}^{(4)}(A)$ & $32.7 \pm 2.1$ & $0.59 \pm 0.02$ & $\,\,\,\,12.2 \pm 0.8\,\,\,\,$ & $0.36 \pm 0.02\,\,\,\,$\\

$\Delta_{np}^{MN}(A)$ & $19.9 \pm1.6$ & $0.48 \pm 0.02$ & $\,\,\,\,15.2 \pm 0.6\,\,\,\,$ & $0.41 \pm 0.01\,\,\,\,$\\
\bottomrule
\end{tabular*}
\end{center}
\vspace{20pt}


\begin{multicols}{2}

Diagrams for the indicators $\Delta_{np}^{MN}$ \eqref{D13} and $\delta_{np}$ \eqref{dnp} are shown in Fig.~\ref{pic:  Recipes}\textit{i} and \textit{f}. From the diagrams, the relationship between $\delta_{np}$ and previously introduced indicators $\Delta_{np}$, $\Delta^{(6, n)}_{np}$ and $\Delta^{(6, p)}_{np}$ is clear: in fact, as in the case of identical nucleons, the relations from  \cite{MN} are a further averaging of the $np$-interaction energy $\Delta_{np}$ over the mass surface.

As it was mentioned above, OES indicator $\Delta^{MN}_{np}$~\eqref{D13} has a very inderect relation to $np$-correlation, but we included it into consideration as a well-studied reference point.
According to the scheme in Fig.~\ref {pic:  Recipes} $ i) $, one can see that this relation $\Delta^{MN}_{np}$ for $A$-even nuclei is also an averaging, but of indicator $\Delta^{(3)}_{np} $~\eqref{d3}:
\begin{align}
\nonumber \Delta_{np}^{MN}(N,Z)=\frac{1}{4}&\left(2 \Delta^{(3)}_{np}(N,Z) -\Delta^{(3)}_{np}(N+1,Z-1) - \right. \\
&\left.-  \Delta^{(3)}_{np}(N-1,Z+1)\right), 
\end{align}

It is interesting to note some similarities in the construction of $\Delta^{MN}_{np}$~\eqref{D13} and expression of $d$-term in Wigner energy $d(A)$ (\ref{Wign}). The latter can also be represented as $\Delta^{(3)}_{np} $ combination:
\begin{align}
\nonumber d(A)=\frac{1}{2}&\left( \Delta^{(3)}_{np}(N,Z) +\Delta^{(3)}_{np}(N+2,Z-2) - \right. \\
&\left.-  \Delta^{(3)}_{np}(N,Z-2)- \Delta^{(3)}_{np}(N+2,Z)\right). 
\end{align}

The values $\Delta_{np}^{MN}(A)$ of  (\ref{D13}) for nuclei with $ N = Z $ are shown in Fig.~\ref {dnp},~b) in comparison with $\Delta_{np} (A) $. The dependence $\Delta_{np}^{MN} (A)$ is smoother, but in the region of nuclei with $ A> 40 $ all three indicators coincide with good accuracy.

Table~\ref{Cb} shows the results of fitting dependencies $\Delta_{np}(A)$, calculated using the formulas discussed earlier, for the chain of nuclei with $N = Z$ using the power function $\Delta_ {np}( A) = C \cdot A^{-b}$. In general, the results can be divided into two large groups. Indicators, appropriate assessment of mass splitting $ \Delta_{np}^{MN}$, $\Delta_{np}^{(3)}$ and $\Delta_{np}^{(4)} $ and estimation based on the definition of $np$-pairing in odd-odd nuclei, $\Delta_{np} $ and $\Delta_{np}^{(7)}$ correspond to approximation \cite{BM} $2\Delta = 24/A^{1/2} $, and the exponent  can be approximated with sufficient accuracy by the power functions $ A^{1/2} $ or $ A^{2/3} $ used to describe the pairing energy of nucleons in modern macroscopic models. Coefficient of the neutron OES effect fitting by $\Delta_n^{(4)} = C_n \cdot A^{-1/2} $ at the current data set is slightly less than $12$~MeV, $C_n = 10.77 \pm 0.06 $~MeV \cite{VMU14}.  This result is in best agreement with the fitting parameters for $\Delta_{np}^{MN}(A)$. Such an outcome can be explained by the fact that the smoothest formulas were used for approximation of both $\Delta_n(A)$ and $\Delta_{np}(A)$.

Significantly smaller values on the whole range of nuclei correspond to the pairing energy formulas $\Delta_{np}^{(6n)} $, $\Delta_{np}^{(6p)} $ and $\delta_{np} $. Small values in combination with significant fluctuations indicate the unreliability of the approximations of these quantities. It should be noted that the coefficients of the approximations of these characteristics are in good agreement with each other.

\section{Shell model}\label{Shells}

The first step in interpreting the mass relations obtained can be made within the framework of the shell model \cite{Z58}. Consider a nucleus with $n$ neutrons in the state $ j_1$ and $p$ by protons in the state $j_2$ above the closed core $(N_0,Z_0)$. The binding energy of such a configuration can be represented as a sum:
\begin{align}
\nonumber B(N_0+n,Z_0+p)=B(N_0,Z_0)+n\varepsilon_n+p\varepsilon_p+\\
+W(j^n_1)+W(j^p_2)+I(j_1^n,j_2^p) ,
\label{SM1}
\end{align}
where $\varepsilon_n$ and $\varepsilon_p$ denotes single-particle  central-field energies of the $j_1$ neutrons and $j_2$ protons, respectively. Terms  $ W (j)$ correspond to the interaction energy of nucleons in a given shell, $ I (j_1,j_2)$ denotes the energy of interaction between nucleons located on different shells.
The contribution of the interaction of $n$ identical nucleons in the state $ j $ can be written as the sum of two terms:
\begin{equation}
W(j^n)=\frac 12 \left(n-\frac{1-(-1)^n}{2}\right)\pi+\frac{n(n-1)}{2}d ,
\end{equation}
the first of which is responsible for the coupling of identical nucleons with "pairing energy" $\pi$. The second term describes additional interaction of two nucleons with strength $d$, independent of relative orientation of their spins and having the character of repulsion.  Ratio of these quantities is clearly seen in the dependence $S_n(N)$ for $Z =$~Const (Fig.~\ref {pic: Sn(N)}). The pairing energy $ \pi $ is responsible for the zigzag behaviour of the curve and determined by the difference between $S_n(N) $ in neighbouring nuclei with even and odd $N$. The value of $d$ assigns the total slope of the curve and can be estimated via the difference $S_n(N + 1)-S_n (N-1) $. Fig.~\ref{pic: D6p} shows a scheme that allows one to estimate the values of $\pi$ and $d$ on the basis of  $S_n(N) $  in isotones. The mass difference relations for identical nucleons were considered in our previous paper \cite{CPC} in detail.

The interaction of $n$ neutrons in $j_1$ state and $p$ protons in $j_2$ state can be written as the sum of two terms  \cite{RT52}:
\begin{equation}
I(j_1^n,j_2^p)=npI^0+\frac{(1-(-1)^n)(1-(-1)^p)}{4}I' ,
\end{equation}
where the contribution $I^0$ does not depend on the nucleon spin orientation and is determined by scalar interaction, while the contribution of $ I '$  depends on the value of the total spin $J$, represents the "pairing properties"  of interaction  and, accordingly, is present  in odd-odd nuclei only.

Thus, the relation (\ref{SM1}) can be rewritten in the form \cite{Z58}:
\begin{align}
\nonumber  & B(N_0+n,Z_0+p)= \\
\nonumber =&B(N_0,Z_0)+n\varepsilon_n+p\varepsilon_p+\frac{n}{2}\pi_n+ \frac{p}{2}\pi_p+\\
& +\frac{n(n-1)}{2}d_n +\frac{p(p-1)}{2}d_p+npI^0-\delta,
\label{SM2}
\end{align}
where the parity term $\delta$ is given by
\begin{align}
\delta =
\begin{cases}
0, &  ee,\\
\frac{1}{2}\pi_p, & eo,\\
\frac{1}{2}\pi_n, & oe,\\
\frac{1}{2}\pi_n+\frac{1}{2}\pi_p-I', & oo.
\end{cases}
 \label{dsm}
\end{align}
This ratio is simplistic but allows one to identify some regularities in the behaviour of the indicators, based on mass differences.

The neutron separation energy in this representation depends on the parity of $N$ and $Z$:
\begin{align}
S_{n}(N,Z) =
\begin{cases}
\varepsilon_n+(n-1)d_n+pI^0+\pi_n, &ee\\
\varepsilon_n+(n-1)d_n+pI^0,            &oe\\
\varepsilon_n+(n-1)d_n+pI^0+\pi_n - I', &eo\\
\varepsilon_n+(n-1)d_n+pI^0+I', & oo
\end{cases}
 \label{Sn_sh}
\end{align}

Then for pairing of neutrons in an even-even nucleus the following relations hold:
\begin{align}
\Delta_{nn} & = \pi_n+d_n, \label{3}\\
\Delta_{nn}^{(3)} & = \pi_n-d_n, \label{4} \\
\Delta_{nn}^{(5)} & =2\Delta_n^{(5)} = \pi_n.
\end{align}

In terms of this model, since $d <0$, $\Delta_{nn} $ in case of even $N$ is always less than for odd $N$,  quantity $\Delta_{nn}^{(3)}$ has an inverse relation, and the averaging characteristic $\Delta^{(5)}_{nn}$ corresponds only to neutron pairing energy $\pi_n$. The value of $d$ can also be extracted from the mass relations data, but as the difference $ (\Delta_{nn} - \Delta_{nn}^{(3)})/2$.

\end{multicols}

\ruleup
\begin{center}
\includegraphics[width=12cm]{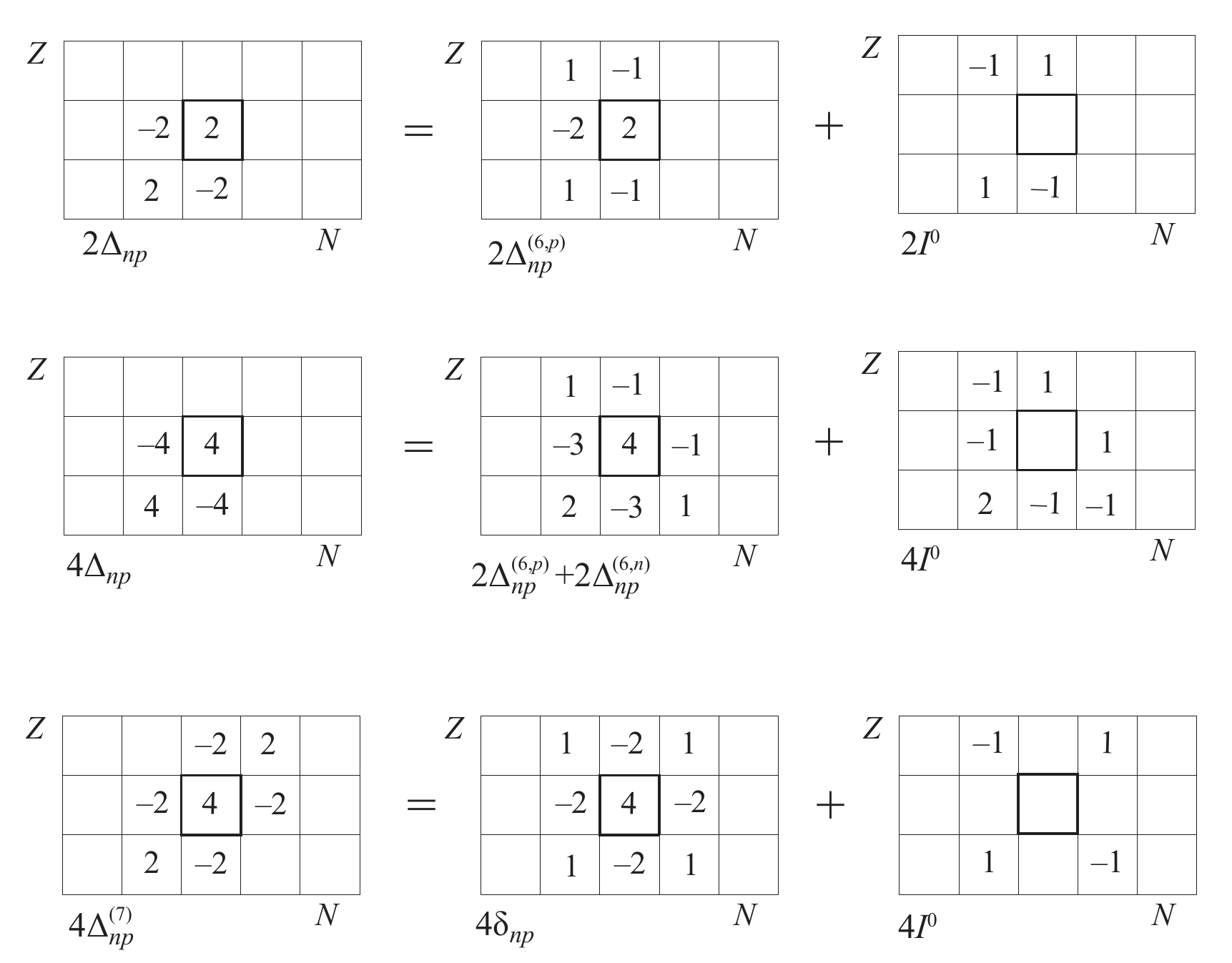}
\figcaption{Diagrams for estimation of energy of $np$-pairing $\Delta_{np}$, as well as $ I^0$ and $I '$. See text for details}
\label{pic:II nod}
\end{center}

\ruledown

\begin{multicols}{2}

\subsection{Neutron - proton interaction}

Let us consider the structure of the previously introduced mass relations for $np$-correlations. The values of indicators $\Delta_{np}$ (\ref{Def_np}) and $\Delta_{np}^{(7)}$ (\ref{Delta_7})) significantly differ for even and odd $A$:
 \begin{align}
\Delta_{np} = \Delta_{np}^{(7)} = I'+I^0 & ( ee,oo), \label{npe}\\
\Delta_{np} = \Delta_{np}^{(7)} = I'-I^0  & (oe, eo).
 \label{npo}
\end{align}

Earlier, a good agreement of these relations was pointed out on the example of even $A$ nuclei (see Fig.~\ref{pic: dVnp}). It should be noted here that averaging undertaken in the relation for $\Delta_{np}^{(7)}$ does not allow one to separate contributions of $I^0$ and $I'$. 

The contribution of $I'$ is mapped, by analogy with  like nucleon pairing, by indicators $\Delta_{np}^{(6,n)}$ (\ref{Delta_6n}) and $\Delta_{np}^{(6,p)}$ (\ref{Delta_6p}):
$$\Delta_{np}^{(6,n)} = \Delta_{np}^{(6,p)} = I' $$
and consequently the indicator $\delta_{np} (N,Z)= (\Delta_{np}^{(6,p)}(N,Z)+\Delta_{np}^{(6,p)}(N+1,Z))/2= I'$ (see (\ref{dnp})).  Comparison of the diagrams for indicators $\Delta_{np}$ and  $\Delta_{np}^{(6,p)}$ (see fig.~\ref{pic:II nod}, first row) leads to the expression for the parameter $I^0$ in even-$A$ nuclei: 
\begin{align}
\nonumber I^0(N,Z) =& \frac{1}{2}[B(N,Z+1)-B(N-1,Z+1)+\\
&+B(N-1,Z-1)-B(N,Z-1)].
\end{align}

One can obtain a similar formula  with $\Delta_{np}^{(6,n)}$. Since the results of $\Delta_{np}^{(6,n)}$ (\ref{Delta_6n}) and $\Delta_{np}^{(6,p)}$ (\ref{Delta_6p}) calculations differ slightly as was shown earlier, it is efficient to average them. 

\begin{center}
\includegraphics[width=7.5cm]{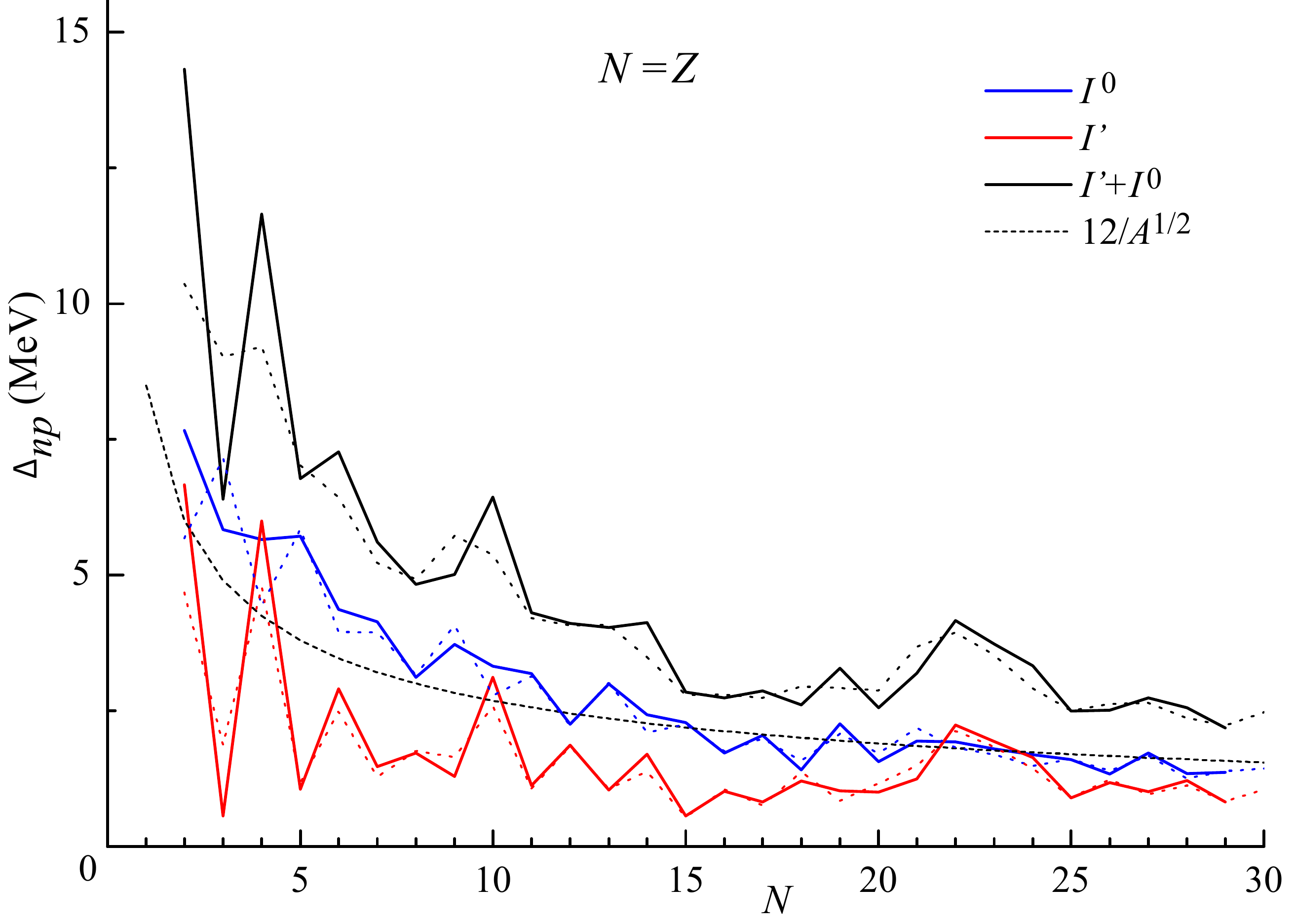}
\includegraphics[width=7.5cm]{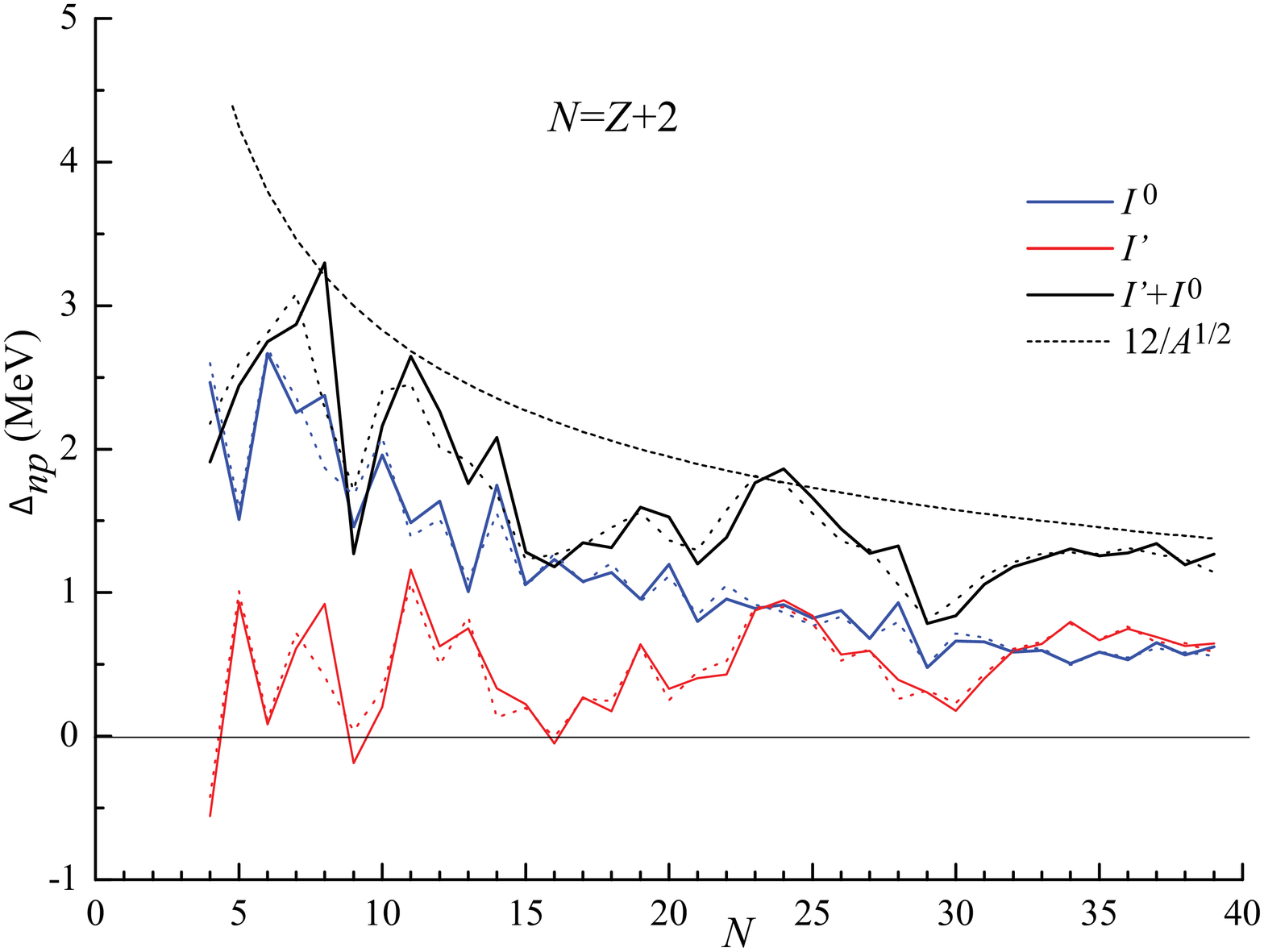}
\figcaption{Comparison of $\Delta_{np}, I^0, I'$ in chains of nuclei with 
 $N=Z$ (a) and $N-Z=2$ (b). Solid lines present results for averaged parameters from second row on Fig.~\ref{pic:II nod} ($I'=(\Delta_{np}^{6,p}+\Delta_{np}^{6,n})/2$), dashed lines correspond to the third row ($I'=\delta_{np}$).}
\label{I0I}
\end{center}

\begin{center}
\tabcaption{\label{CbII}  Coefficients $C$ (MeV) in $\pi$ and  $d$ for like nucleons approximation by the power function  $C\cdot A^{-b}$}
\begin{tabular*}{80mm}{c@{\extracolsep{\fill}}ccc}
\toprule 
  & Neutrons &  Protons  \\
\hline
$\pi(A)=C/A^{1/3}\,\,\,\,$ & $\,\,\,\,10.22 \pm 0.06\,\,\,\,$ &  $\,\,\,\,11.48 \pm 0.06\,\,\,\,$ \\
$d(A)=C/A\,\,\,\,$ & $\,\,\,\,-23.0 \pm 0.3\,\,\,\,$ & $\,\,\,\,-56.7 \pm 0.6 \,\,\,\,$ \\
\bottomrule
\end{tabular*}
\vspace{0mm}
\end{center}

Diagrams for this case are shown on the second row in Fig.~\ref{pic:II nod} and calculations in accordance with these diagrams are presented in Fig.~\ref{I0I} by solid lines.  The values of $I^0$ lay above $I'$ and for nuclei with $N=Z$ well correspond to the dependence $12/A^{1/2} $, which was proposed to describe the pairing effect. The values of $I'$ fluctuate much more strongly.  With increasing $A$, the values of the parameters become closer and the ratio between them can vary. Dotted line shows parameters $I^0$ and $I'$, calculated by the most averaged formula for $I' = \delta_{np}$. The diagrams for calculating the parameters in this case have the most symmetrical form (see the third row of Fig.~\ref{I0I}). In this case, the sum of the parameters $I^0+I'$ is endorsed by the indicator $\Delta_{np}^{(7)}$, and the expression for the parameter $I^0$ is of the form:
\begin{align}
\nonumber I^0(N,Z) =& \frac{1}{4}[B(N+1,Z+1)-B(N-1,Z+1)+\\
&+B(N-1,Z-1)-B(N+1,Z-1)].
\end{align}

This formula coincides with the expression for empirical $np$-interaction of the last neutron with the last proton in even-even nuclei $\delta V_{np}$ from \cite{JC85, ZCB89, BW90}.   The important point is that in this case the binding energies of odd-odd nuclei are utilized in calculation, in contrast to relation (\ref{Def_ee}), where the calculation of $\delta V_{np}^{ee}$ is based on $B(N,Z)$ in even-even isotopes. This difference does not lead to significant changes in the numerical results in general approximation constructions, but, apparently, should be taken into account in more accurate model descriptions.

Table~\ref{CbII} presents the coefficients of approximation of parameters $\pi$ and $d$ for the identical nucleon interaction by the function $C\cdot A^{-b}$ (MeV). Parameters were fitted without taking into account magic and self-adjoint nuclei in accordance with the selection rules from  \cite{Moller_Nix}. The values of the fixed exponent $b$ were chosen to be the closest to the results of approximation with two free parameters: $C$ and $b$. For the parameters of  pairing of neutrons $\pi_n$ and protons $\pi_p$, the values of $b$ were $ 0.30 \pm 0.01 $ and $ 0.32 \pm 0.01 $, which is close to 1/3. For  parameters $d_n$ and $d_p$, the value of $b$ was chosen equal to unity, which agrees well with the value of the selected coefficient $b$ for neutrons ($0.93 \pm 0.03$). In the case of protons, the deviation of the adjusted coefficient $b$ from unity is more significant due to the effect of Coulomb interaction ($0.56 \pm 0.01$). 

\begin{center}
\tabcaption{\label{CbIII} Coefficients $C$ (MeV) in $ I^0 $ and $ I'$ approximation by the power function $ C \cdot A^{-b} $ for various fixed $b$}
\begin{tabular*}{80mm}{c@{\extracolsep{\fill}}ccc}
\toprule 
  & $I^0$ &  $I'$  \\
 \hline
$C/A\,\,\,\,$ & $\,\,\,\,41.9 \pm 0.3\,\,\,\,$ & $\,\,\,\,30.7 \pm 0.4\,\,\,\,$ \\
$C/A^{2/3}\,\,\,\,$ & $\,\,\,\,9.43 \pm 0.06\,\,\,\,$ & $\,\,\,\,7.04 \pm 0.08\,\,\,\,$ \\
$C/A^{1/3}\,\,\,\,$ & $\,\,\,\,1.93 \pm 0.02\,\,\,\,$ & $\,\,\,\,1.46 \pm 0.02\,\,\,\,$ \\
\bottomrule
\end{tabular*}
\vspace{0mm}
\end{center}

The values of the coefficient $C$ in $I^0$ and $I'$ approximation by the power function $C \cdot A^{-b}$ for various values of $b $ are presented in table~\ref{CbIII}. The fixed values of $b$ allow us to compare parameters $I^0 $ and $I'$ to each other, and also to compare them with $\pi$ and $d$ values. Approximation by a power-law function with a free exponent gives the values of $b$ equal to $0.83 \pm 0.01$ and $0.67 \pm 0.02$ for parameters $I^0$ and $I'$ respectively. Thus, for $I^0$ the best approximation is the $C/A$ dependence, whereas the $C/A^{2/3}$ ratio should well correspond to all values of $I'$. For all variants of $b$, the coefficients $C$ for $I^0$ and $I'$ have close values, and always $I^0 > I'$.

\subsection{"Deuteron-type" relations and mass staggering}

The splitting of the mass surface $\Delta^{MN}_{np}(N,Z)$, determined by  formula~(\ref{D13}),  does not depend on parity of $N$ and $Z$ in the shell model approach with parameterization~(\ref{SM1}) and has the form:
\begin{equation}
\Delta^{MN}_{np} (N,Z)=\frac{\pi_n}{2}+\frac{\pi_p}{2}-I' ,
\end{equation}
which corresponds to the definition of given indicator. However, it is important to note that this relation is written just for nuclei with even $A$, and in application to odd nuclei the meaning of this characteristic is not obvious. More indicative are the values $\Delta_{np}^{(3)}$ and  $\Delta_{np}^{(4)}$. Formula for $\Delta_{np}^{(3)}(N,Z)$ depends on $N$ and $Z$ parity:
\begin{align}
\nonumber&\Delta_{np}^{(3)}(N,Z) =\\
&=\frac12
\begin{cases}
\left((\pi_n - d_n)+(\pi_p - d_p)\right)-2(I'+I^0), &  ee\\
\left(-(\pi_n - d_n)+(\pi_p + d_p)\right)+2I^0,  & oe\\
\left((\pi_n + d_n)-(\pi_p - d_p)\right)+2I^0, &  eo\\
\left((\pi_n + d_n)+(\pi_p + d_p)\right)-2(I'-I^0), & oo
\end{cases}
\label{D3_sh}
\end{align}
The expression for even-even nuclei corresponds to splitting of the mass surface between even-even and odd-odd nuclei
$$\Delta_{np}^{(3)}(ee)=\frac{1}{2}\left(\Delta_{nn}^{(3)}(ee)+\Delta_{pp}^{(3)}(ee)\right)-\Delta_{np}(ee),$$
the ratios for nuclei with odd $A$ contain the energy difference of identical nucleons pairing and correspond to a small splitting of the mass surface between even-odd and odd-even nuclei.

The relations for indicator $ \Delta_{np}^{(4)}$  depend on the parity of $A$:
\begin{align}
\Delta_{np}^{(4)}(N,Z) = \frac12
\begin{cases}
\left(\pi_n +\pi_p \right)-2I', &  ee, oo\\
\left( d_n+ d_p\right)+2I^0,            &  oe, eo
\end{cases}
 \label{D4_sh}
\end{align}
This expression for even $A$ coincides with the expression for $\Delta^{MN}_{np}$. The degree of this equality can be seen in Fig.~\ref{pic:dnp} through the example of a chain of $N=Z$ nuclei. The figure shows the dependencies of indicators $\Delta^{MN}_{np}$, $\Delta_{np}^{(3)}$ and  $\Delta_{np}^{(4)}$ on the mass number and it is clear that while $\Delta^{MN}_{np}$ and $\Delta_{np}^{(4)}$  coincide well only in the region $A>40$, indicators $\Delta_{np}^{(3)}$ and $\Delta_{np}^{(4)}$ coincide with good accuracy for all  $N$ and $Z$, except for the values of magic numbers. From the approximate equality $\Delta_{np}^{(3)} \approx \Delta_{np}^{(4)}$  for odd-odd nuclei follows:
\begin{align}
\nonumber \frac{1}{2}\left((\pi_n + d_n)+(\pi_p + d_p)\right)-(I'-I^0) &\approx \frac{1}{2}\left(\pi_n +\pi_p \right)-I',\\
\frac{1}{2}\left( d_n+ d_p\right)+I^0& \approx 0
\end{align}

The last relation connects the values of parameters $d$ and $I^0$, and also asserts that for odd $A$, the values $\Delta_{np}^{(3)}$ and $\Delta_{np}^{(4)}$ are close to zero. Indeed,  the estimates of $\pi$, $d$ and $I$, made in \cite{Z58} on the array of stable nuclei, shows that parameter $d_n$ is about $-0.1$~MeV,  parameter $d_p$ is about $-0.5$~MeV, and $I^0$ has a value of about $0.3$~MeV. Proximity to zero of values of $\Delta_{np}^{(3)}$ and $\Delta_{np}^{(4)}$ for odd $A$ indicates the equality of pairing forces of identical nucleons:
\begin{align} 
\nonumber \frac{1}{2}\left(-(\pi_n - d_n)+(\pi_p + d_p)\right)+I^0 &\approx \frac{1}{2}\left( d_n+ d_p\right)+I^0 \approx 0,\\
\pi_p& \approx \pi_n,
\end{align}
which follows from the charge independence of nuclear forces. The degree of fulfillment of these relations is clearly seen in the values of the coefficient $C$ of  $\pi$, $d$ and $I$ approximations given in the tables~\ref{CbII} and \ref{CbIII}.

\section{Conclusions}

Mass relations based on  even-odd staggering of the mass surface are widely used to estimate the identical nucleon pairing in an atomic nucleus. By analogy, a significant number of mass indicators are constructed for $np$-correlations in order estimate the value of the $np$-pairing. However, the difficulty of extracting experimental information for odd-odd nuclei significantly limits the possibilities for analyzing the values obtained.

In this paper, various indicators $\Delta_{np}$ are considered by the example of both odd-odd and even-even nuclei.  Both estimates of the $np$-pairs separation and the relationships constructed by analogy with the estimation of like nucleon pairing, can serve as the basis for construction of mass ratios. It turns out that most of the ratios are related to one another and are based on the basic assessment of neutron-proton correlation in  odd-odd nucleus:
\begin{align}
\nonumber \Delta_{np} =& B(N,Z)+B(N-1,Z-1)-\\
\nonumber &-B(N-1,Z)-B(N,Z-1),
\end{align}
representing  either an average of the given characteristic, or a difference of its values for the neighbouring nuclei.
Thus, the widely discussed characteristic $\delta V_{np}$ coinciding with the definition of $\Delta_{np}$ for  odd-odd nuclei, represents an average of $\Delta_{np}$ over four isotopes when applied to even-even nuclei. The correction to the $np$ interaction $\delta_{np}$, commonly mentioned in the discussion of the mass surface splitting, also serves as averaging of $\Delta_{np}$ over four neighbouring nuclei, but performed taking into account the zigzag feature of the dependence of the neutron separation energy in isotopes (or proton separation energy in isotones).

This approach is similar to the method used to get an estimate of like nucleon pairing energy, but it can bring us to essentially different results. In general,  the zigzag relation  depends on two parameters $\pi$ and $d$, which fix the amplitude of the oscillations and the general slope of the dependence. While for identical nucleons, where the mass relations are constructed on the basis of $S_n(N)$ and $S_p(Z)$ dependencies in isotopes and isotones respectively, the values $\pi$ and $d$ differ by several times,  the corresponding quantities for dependencies $S_n(Z)$ in isotones and $S_p(N)$ in isotopes are close in magnitude.  Furthermore, the relationship between various characteristics of nuclei within a single isotope chain, can change. Such changes affect inevitably the results of calculations by formulas analogous to relations for like nucleon pairing, and lead to appearance of alternating quantities.

To clear up the structure of various indicators $\Delta_{np}$,  parametrization of the binding energy of the atomic nucleus based on the shell model, was used. This approach effectively takes into account both the residual interaction of identical nucleons in one state and the interaction of nucleons on different subshells, such as external neutrons and protons between themselves. Such a parametrization makes it possible to show the interrelation of different mass ratios more clearly and  to elucidate their physical meaning. In context of this parametrization, the pair interaction of identical nucleons is described by the sum of two terms
$$\Delta_{nn(pp)} = \pi_{n(p)} + d_{n(p)}.$$
The first of which is responsible for pairing of identical nucleons with "pairing energy" $\pi$, while the second term describes additional interaction of nucleon pair with strength $d$, independent of relative orientation of the nucleon spins and having a repulsive character. However, taking into account the ratio of $\pi$ and $d$, using one parameter $\pi$ to describe pairing forces does not greatly affect the result. In this approach, the $np$ interaction in odd-odd nuclei should include both contributions
$$\Delta_{np} = I^0+I',$$
and consequently due to proximity of the quantities $I^0$ and $I'$, accounting for only one parameter changes the result by two times. This is most clearly seen from the comparison of  indicator $\Delta_{np}$ with averaging characteristics $\Delta_{np}^{(6,n)}$, $\Delta_{np}^{(6,p)}$ and $\delta_{np}$, the values of which are about $\Delta_{np}/2$.

Approximations of parameters $I^0$ and $I'$ by the power function $C/A^b$ with various fixed values of $b$ allows us to demonstrate a clear relationship of various parameters on the whole modern array of atomic nuclei. Thus, coefficients $C$ of $ d_n $, $ d_p $ and $ I^0 $ approximations by the dependence $ C/A $ are of the same order of magnitude and  are approximately tied by relation $d_n+d_p \approx -2I^0$.  In turn, the pairing parameters of identical nucleons $ \pi_n $ and $ \pi_p $ are well described by the $ C/A^{1/3} $ dependence for close values of the coefficients $ C $ above 10~MeV. The value of the coefficient $ C $ in the $ I '$ approximation by the $C/A^{1/3}$ dependence is almost an order of magnitude smaller, $1.38 \pm 0.02 $~MeV, which clearly illustrates the relation of the pairing effects of identical nucleons and $np$-interaction.

 \acknowledgments{The authors would like to thank Dr. D.~Lanskoy and L.~Imasheva for fruitful discussions and technical support. The work of  S. Sidorov was supported by Foundation for the advancement of theoretical physics and mathematics BASIS.}

\end{multicols}

\vspace{-1mm}
\centerline{\rule{80mm}{0.1pt}}
\vspace{2mm}

\begin{multicols}{2}

\end{multicols}

\clearpage
\end{document}